%% file: main.tex
\documentclass[sn-mathphys]{sn-jnl}% Math and Physical Sciences Reference Style
\usepackage[normalem]{ulem}  
\newcommand\redout{\bgroup\markoverwith
{\textcolor{red}{\rule[0.5ex]{2pt}{0.8pt}}}\ULon}

\newcommand{\note}[1]{\textcolor{black}{[#1]}}

\jyear{2023}%

\theoremstyle{thmstyleone}%
\theoremstyle{thmstyletwo}%

\theoremstyle{thmstylethree}%

\begin{document}

\title[BalanceVR]{BalanceVR: 
Balance Training to Increase Tolerance to Cybersickness in Immersive Virtual Reality}

%%=============================================================%%
%% Prefix	-> \pfx{Dr}
%% GivenName	-> \fnm{Joergen W.}
%% Particle	-> \spfx{van der} -> surname prefix
%% FamilyName	-> \sur{Ploeg}
%% Suffix	-> \sfx{IV}
%% NatureName	-> \tanm{Poet Laureate} -> Title after name
%% Degrees	-> \dgr{MSc, PhD}
%% \author*[1,2]{\pfx{Dr} \fnm{Joergen W.} \spfx{van der} \sur{Ploeg} \sfx{IV} \tanm{Poet Laureate} 
%%                 \dgr{MSc, PhD}}\email{iauthor@gmail.com}
%%=============================================================%%

\author[1]{\fnm{Seonghoon} \sur{Kang}}\email{econksh@gmail.com}
\equalcont{These authors contributed equally to this work.}

\author[1]{\fnm{Yechan} \sur{Yang}}\email{santachan@korea.ac.kr}
\equalcont{These authors contributed equally to this work.}

\author[1]{\fnm{Gerard Jounhyun} \sur{Kim}}\email{gjkim@korea.ac.kr}
%\equalcont{These authors contributed equally to this work.}

\author[1,2]{\fnm{Hanseob} \sur{Kim}}\email{khseob0715@korea.ac.kr}

\affil[1]{\orgdiv{Computer Science and Engineering}, \orgname{Korea University}, \orgaddress{\street{145, Anam-ro, Seongbuk-gu}, \city{Seoul}, \postcode{02841}, \state{Seoul}, \country{South Korea}}}

\affil[2]{\orgdiv{Center for Artificial Intelligence}, \orgname{Korea Institute of Science and Technology}, \orgaddress{\street{5, Hwarang-ro 14-gil, Seongbuk-gu}, \city{Seoul}, \postcode{02792}, \state{Seoul}, \country{South Korea}}}

%\affil[3]{\orgdiv{Department}, \orgname{Organization}, \orgaddress{\street{Street}, \city{City}, \postcode{610101}, \state{State}, \country{Country}}}

%%==================================%%
%% sample for unstructured abstract %%
%%==================================%%
% \vspace{-0.3em}

\abstract{
Cybersickness is a serious usability problem in virtual reality.  
Postural (or balance) instability theory has emerged as one of the primary hypotheses for the cause of VR sickness.  
In this paper, we conducted a two-week-long experiment to observe the trends in user balance learning and sickness tolerance under different experimental conditions to analyze the potential inter-relationship between them.  
The experimental results have shown, aside from the obvious improvement in balance performance itself, that accompanying balance training had a stronger effect of increasing tolerance to cybersickness than mere exposure to VR.  
In addition, training in VR was found to be more effective than using the 2D-based non-immersive medium, especially for the transfer effect to other non-training VR content.}


%%\pacs[JEL Classification]{D8, H51}

%%\pacs[MSC Classification]{35A01, 65L10, 65L12, 65L20, 65L70}

\maketitle

\input{sections/01_intro}
\input{sections/02_relatedwork}

\input{sections/03_Experiment}
\input{sections/04_Results}

\input{sections/06_Discussion}
\input{sections/07_Conclusion}

\backmatter

% \bmhead{Supplementary information}
% If your article has accompanying supplementary file/s please state so here.

\bmhead{Acknowledgments}
This research was supported by the Basic Research Laboratory Program funded by NRF Korea (2022R1A4A1018869), the ITRC Program (IITP-2022-RS-2022-00156354) funded by MSIT/IITP Korea, and the Competency Development Program for Industry Specialist funded by MSIT/IITP/MOTIE/KIAT Korea (N0009999)

% \subsubsection*{}

% \subsubsection*{Ethics approval }

% \begin{itemize}
% % \item Funding 
% \item Conflict of interest

% % /Competing interests (check journal-specific guidelines for which heading to use)
% % \item Ethics approval 
% % Thiss
% % \item Consent to participate
% % \item Consent for publication
% % \item Availability of data and materials
% % \item Code availability 
% % \item Authors' contributions
% \end{itemize}

%\noindent
%If any of the sections are not relevant to your manuscript, please include the heading and write `Not applicable' for that section. 

%%===================================================%%
%% For presentation purpose, we have included        %%
%% \bigskip command. please ignore this.             %%
%%===================================================%%

\begin{appendices}

\input{sections/A0_Appendix}

%%=============================================%%
%% For submissions to Nature Portfolio Journals %%
%% please use the heading ``Extended Data''.   %%
%%=============================================%%

%%=============================================================%%
%% Sample for another appendix section			       %%
%%=============================================================%%

%% \section{Example of another appendix section}\label{secA2}%
%% Appendices may be used for helpful, supporting or essential material that would otherwise 
%% clutter, break up or be distracting to the text. Appendices can consist of sections, figures, 
%% tables and equations etc.

\end{appendices}

%%===========================================================================================%%
%% If you are submitting to one of the Nature Portfolio journals, using the eJP submission   %%
%% system, please include the references within the manuscript file itself. You may do this  %%
%% by copying the reference list from your .bbl file, paste it into the main manuscript .tex %%
%% file, and delete the associated \verb+\bibliography+ commands.                            %%
%%===========================================================================================%%
\newpage
\section*{Declarations}

%Some journals require declarations to be submitted in a standardised format. Please check the Instructions for Authors of the journal to which you are submitting to see if you need to complete this section. If yes, your manuscript must contain the following sections under the heading `Declarations':
\begin{itemize}
    \item \textbf{Conflicts of interest} The authors declare that they have no conflict of
interest.
    \item \textbf{Ethics approval} The experiment was approved by the Institutional Review Board (IRB No. 2023-0143-01).
    
    \item \textbf{Availability of data and material} 
    The data generated during and/or analyzed during the current study are available from the corresponding author on reasonable request.

    \item \textbf{Code availability } This code is not available.
\end{itemize}

% \bibliography{sn-bibliography}

%% BioMed_Central_Bib_Style_v1.01

\end{document}

%% file: sections/01_intro.tex
\section{Introduction}

% \cite{chang2020virtual}

The immersive and spatial nature of 3D virtual reality (VR) are important qualities in lending itself as an attractive means for various types of training - including physical exercises~\cite{Lee2018TheEO,Neumann}.
A typical method of physical training would be e.g., understanding and remembering to follow various exercise instructions illustrated on paper, or watching the trainer's motion in the third person viewpoint and mimicking it.
VR may be particularly suited for teaching physical exercises as it can provide the first person viewpoint and convey the spatial/proprioceptive sense of the required movements, e.g., as if enacted by one's own body parts~\cite{Yang,10.1145/3198910.3198917}.
Consequently, many VR-based physical training systems have indeed been developed and shown their effectiveness~\cite{Ahir}.

On the other hand, postural instability has been proposed and emerged 
(although not universally accepted yet) 
as one of the competing hypotheses for the cause of 
cybersickness (or VR sickness)~\cite{Riccio,Behrang,Yoshida}.  
This theory postulates that a VR user can become sick in provocative and unfamiliar 
situations (such as immersed in a VR space)
in which one does not possess (or has not yet learned) strategies or skills to maintain a stable posture and balance~\cite{Riccio,So_Cheung_Chow_Li_Lam_2007, Michael}.
Losing balance is often seen as a consequence (rather than a cause)
of cybersickness, as one major symptom is the dizziness and it has even been used as one
measure for cybersickness~\cite{chang2020virtual, Weech}.  
However, there are also some evidences 
that visually induced sickness (like cybersickness) can be predicted by one's postural instability~\cite{Smart}.
This led us to investigate the VR-based balance training as one possible way 
to increase the tolerance to cybersickness, focusing on a particular and 
clearly observable relevant physical ability.
In this context, it is important that the proposed application of balance training for cybersickness is also situated in the immersive VR environment.  
If shown to be effective, the proposed method can also further corroborate the postural instability theory as well.

In this paper, we analyzed the long-term trends of user balance learning
and sickness tolerance under different experimental conditions and assess the potential inter-relationship between them.  
The transfer effect was also investigated by having the trained users tested for sickness tolerance in non-training VR contents (before and after). 
The main contributions of this paper would be the first-ever findings as follows.

\begin{itemize}
 \item Balance training can be effective in developing the tolerance to sickness from visual motion. 
 \item Immersive VR-based balance training is more effective for developing the tolerance to cybersickness than non-immersive training and mere extended exposure to VR.
 \item The effect of VR-based balance training can be transferred to other VR content 
  (not used for training).
\end{itemize}

% \begin{figure}[t]
% \centering
%   \includegraphics[trim={0 0 0 0}, clip, width=1\textwidth]{figures/firstfigure.jpg}
%   % \vspace{-1em}
%   \caption{Three different cybersickness training methods tested in this paper: 
%   (a) virtual reality-based balance training (VRT), 
%   (b) virtual reality exposure only (VRO), 
%   and (c) 2D projection-based balance training (2DT).
%   On the right, two training VR environments/contents are shown: (d) a jet fighter flight through the forest (Week 1, relatively less sickness), 
%   and (e) a wild roller-coaster ride (Week 2, relatively more sickness).
%   }
%   \label{fig:teaser}
% \end{figure}

\begin{figure}[ht]
\centering
\subfigure[
VRT 
]{\includegraphics[height=14em]{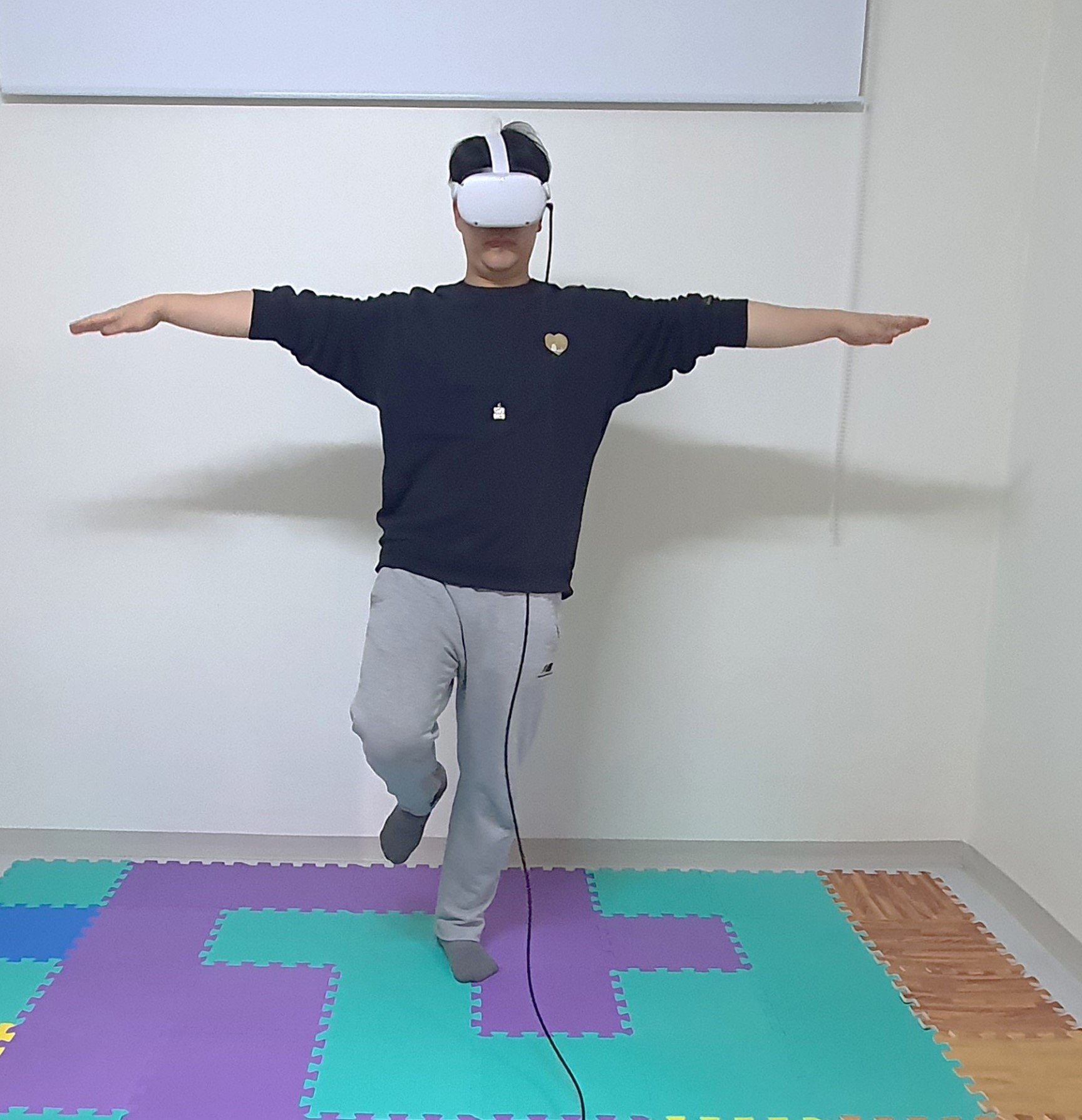}}
\subfigure[
VRO
]{\includegraphics[height=14em]{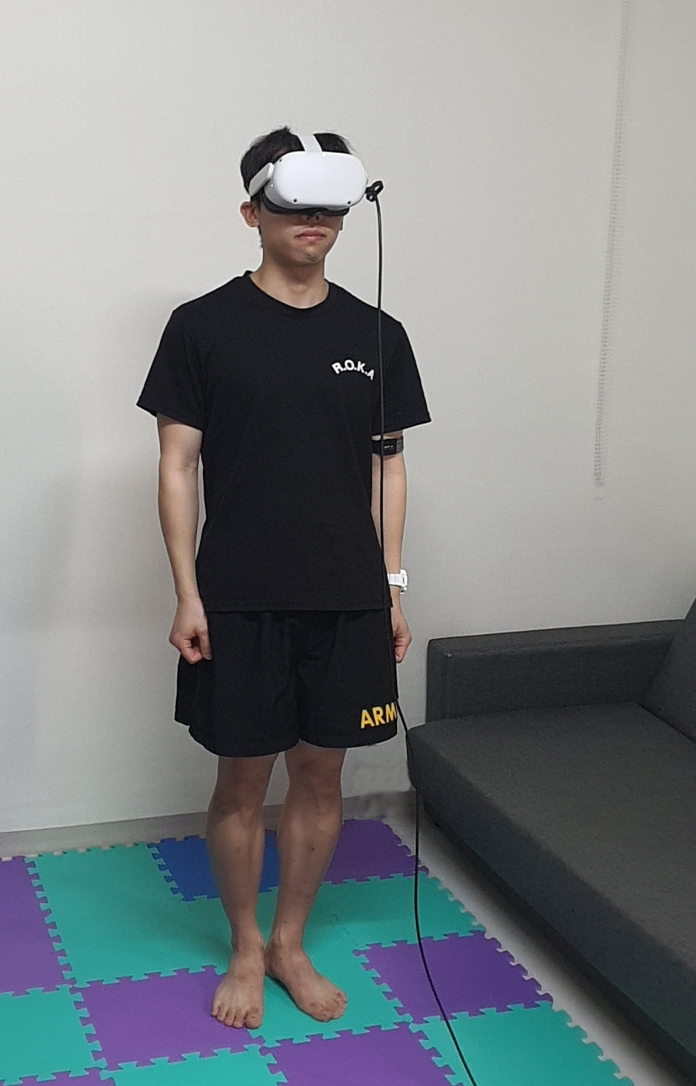}}
\subfigure[
2DT
]{\includegraphics[height=14em]{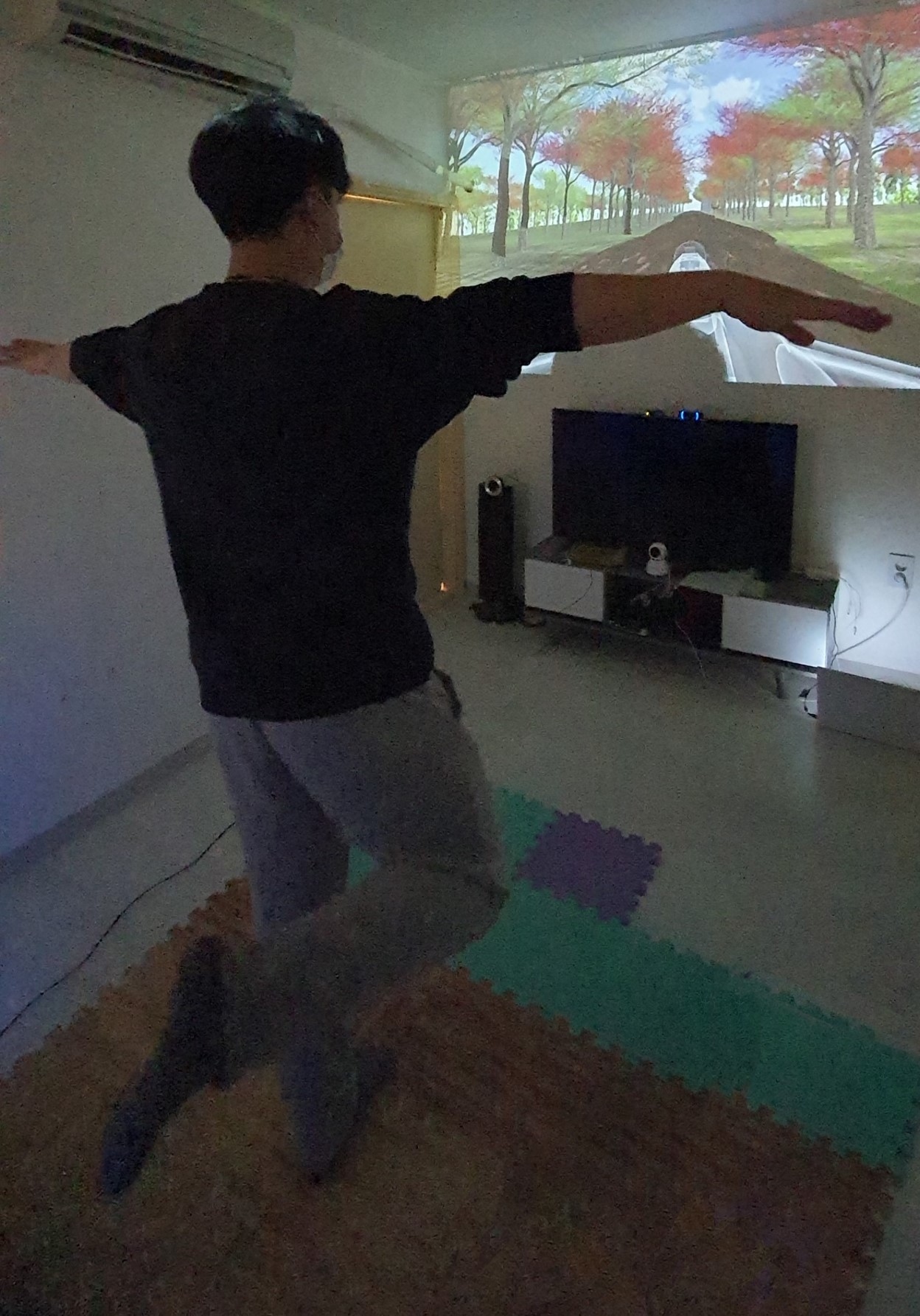}}
\caption{Three different types of cybersickness training methods tested:
(a) Virtual reality-based balance training (VRT);
(b) Virtual reality exposure only (VRO);
and (c) 2D projection-based (non-immersive) balance training (2DT).
The between-subject design was performed.
  \label{fig:teaser}
}
% \vspace{-3em}
\end{figure}

\begin{figure}[ht]
\centering
\subfigure[First week's content
]{\includegraphics[height=9.2em]{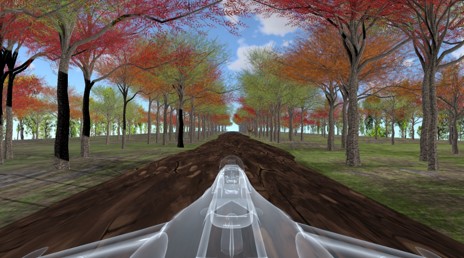}}
\subfigure[Second week's content
]{\includegraphics[height=9.2em]{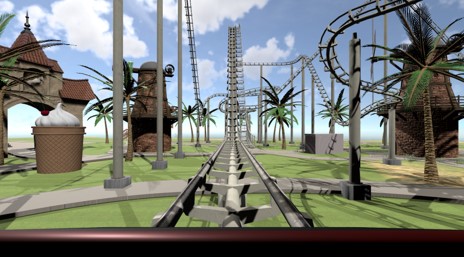}}
\caption{Two training VR environments/contents: (a) a jet fighter flight through the forest (less sickness-inducing) and (b) a wild roller-coaster ride (more sickness-inducing).
  \label{fig:teaser_contents}
}
\end{figure}

%% file: sections/02_relatedwork.tex
\setlength\doublerulesep{0.8pt} 

\section{Related works}

\subsection{Balance Training in VR}

The human body as an articulated and complex skeleton structure is inherently mechanically unstable~\cite{PAI1997347,LEE2006569}.  Maintaining balance (around the center of mass) is a complex process that involves multiple systems in the human body. The vestibular, proprioceptive, and visual channels are used to detect and gather balance/pose-related information and the brain integrates them to coordinate and generate the motor responses and establish the center of pressure through the muscles and joints with constant adjustments to counteract the external perturbation to the body~\cite{PETERKA201827, Lee1998DeterminantsOT, LAFOND20041421}.

There are numerous physical training routines to improve one's balance by strengthening and improving the capabilities of the aforementioned subsystems~\cite{Brachman}.  
Conventional means of balance training usually involve following paper or live instructions from the second/third person point of view.  Virtual reality (VR) can be an effective media offering the first-person perspective 
and sense of personal space in enhancing the understanding of the various training poses and work-outs \cite{10.1145/3198910.3198917, Pastel}.  
Gamification can further provide the motivation and impetus to facilitate the training process \cite{10.1145/3562939.3567818, 10.1145/2909132.2909287}.
However, the effect of balance training (VR-based or not) on cybersickness has not been investigated much despite the fact that they are often touted to be closely related~\cite{Riccio, Haran, 10.1093/ptj/79.10.949}.
% (also see next subsection).

%Regular physical activity, such as strength and balance training, can help improve balance by strengthening muscles, improving proprioception, and enhancing coordination. 

%  이러한 이론을 기반으로 한 균형 트레이닝 연구로는 시각적 자극이 매우 혼란한 상황이나, 눈을 감아서 시각적 자극이 없을 때 어지럼증을 느끼는 전정 미로 손실 환자에게 비슷한 VR 환경을 구성하여, 초점을 고정시키고, 하지의 관절 감각에 집중하여 균형 트레이닝을 진행한 연구가 있었다.[19] 
  
%  그러나 한 명의 노인 환자에게 제공된 연구로 이를 VR 콘텐츠에 대해 일반화한 시도는 아니었다.
  
%  또 다른 VR 환경에서 Balance Training 연구로 가상 환경의 현실 세계를 재구성 할 수 있는 특성을 활용하고 및 게이미피케이션화하여 균형 운동 테스크의 효과성에 대해 조사한 연구가 있었다. 
  
%  그러나 이는 균형 트레이닝 방법에 대한 몇 가지 가능한 VR의 특성을 시도하는 것으로 본 연구에서 검증하고자 하는 사이버 멀미와의 연관성은 없다.[20]

\subsection{Cybersickness}
\label{sec:cybersickness}

Cybersickness, also known as VR sickness, refers to the unpleasant symptoms when using immersive VR simulators, especially with navigational content. 
Typical symptoms include disorientation, headache, nausea, and ocular strains~\cite{laviola2000discussion, kennedy1993simulator}.
The leading explanation for cybersickness is the ``sensory mismatch theory'', which attributes cybersickness to the conflicting user motion information as interpreted by between the visual and vestibular senses~\cite{laviola2000discussion, Rebenitsch}.  
That is, the aforementioned unpleasant symptoms arise when the virtual/visual motion is perceived by the human's visual system while the vestibular senses detect no physical motion. 
Note that the visual and vestibular systems are neurally coupled~\cite{GRSSER1972573}. 
% Several studies have commonly attempted to reduce the amount of or neutralize the visual motion information to reduce the sensory mismatch~\cite{fernandes2016combating, park2022mixing, Keshavarz}.

To combat these symptoms, several studies have focused on reducing the amount of or neutralizing the visual motion information to minimize the sensory mismatch~\cite{fernandes2016combating, park2022mixing, Keshavarz}.
For instance, Fernandes et al.~\cite{fernandes2016combating} developed a dynamic size-shifting field-of-view (FOV) in response to the
speed/angular velocity of users or content.
When the user motion accelerates, the FOV is reduced, which in turn reduces the extent of the visual stimulation and ultimately the sickness. 
% If the velocity is fast, it reduces the FOV size, reducing visual stimulation.
In a similar vein, blurring the peripheral visual field has been proposed to minimize the 
visual stimulation~\cite{budhiraja2017rotation, Lin2020, Caputo2021}.
Park et al.~\cite{park2022mixing, customizedVR} have proposed neutralizing the visual motion stimuli by simultaneously presenting the reverse optical flow.

% \textcolor{red}{Approaches such as 
% are common in attempting to reduce the amount of or neutralize the visual motion information and thereby the extension of the sensory mismatch. 
% }
% \textcolor{blue}{need detail explanation for each method...}

Another popular theory is the rest frame theory, which points to the absence of reference object(s) (objects in the VR content that are not moving with respect to the user), called the rest frame~\cite{Harm}.
The rest frame is thought to help the user maintain one's balance and be aware of the ground (or gravity) direction~\cite{hemmerich2020visually, wienrich2018virtual, Harm}. 
One interesting remedy to cybersickness is the inclusion of the virtual nose, which can be considered as a rest frame object~\cite{wienrich2018virtual, wittinghill2015, cao2017effect}.

Alternatively, a potential strategy for addressing cybersickness might involve methods to alleviate the immediate symptoms and enhance users' physical well-being, rather than directly targeting the root cause. 
These can include e.g. supplying a fresh breeze with a fan ~\cite{igoshina2022comparing, reliefVR} or providing pleasant music or calming aural feedback~\cite{keshavarz2014pleasant, kourtesis2023cybersickness, restingVRPoster}.
%, and reducing the weight of the headset~\cite{reliefVR}.
These measures can be regarded cognitive distraction as a way to reduce the cybersickness by preventing users from focusing on the sickness-inducing VR content~\cite{kourtesis2023cybersickness}.
% 위에서 cognitive distraction에 해당하는 것은 노래를 들려주는 것만인 것 같다는 생각이 듭니다.

One newer
hypothesis, although not fullheartedly accepted in the research community,
for the cause of cybersickness is the postural instability theory~\cite{Riccio, Ruixuan}.  
Accordingly, postural instability, the lack of ability to maintain balance due to 
external factors (such as being subjected to a new unfamiliar, provocative, and thus challenging 
situation) can induce the sickness.  
Note that this does not preclude the fact that imbalance is one typical after-effect of the sickness as well. 
This is based on the various studies that have observed a strong correlation between one's balancing ability (before) and the extent of the motion sickness (after)~\cite{Riccio, Ruixuan, Smart}.
This theory is also in line with the rest frame theory - i.e., the lack of the rest frame object (indicating the direction of gravity and help one maintain balance) could be seen as a provocative situation for the user~\cite{hemmerich2020visually, wienrich2018virtual, Harm}.

Based on all these studies, one can posit that balance training while navigating in the immersive VR would make the user to be even more unstable and exacerbate the extent of the cybersickness. 
In turn, this could make the balance training itself even harder~\cite{imaizumi2020virtual, horlings2009influence}.
Nevertheless, given that the user can endure through the training, its effect can eventually ease and break this vicious cycle.
We can further hypothesize that the immersive feedback will be an important factor, as maintaining and training for balance involves the visual channel and spatial awareness, which non-immersive and 2D-oriented media is difficult to provide fully.

%There have been several research that has used postural instability as a measure to assess the degree of cybersickness~\cite{chang2020virtual, Weech}, 
%especially as an alternative to the survey-based method such as the simulator sickness questionnaire (SSQ)~\cite{kennedy1993simulator}.

On a related note, the length of time exposed to a virtual environment is known to affect the severity of cybersickness~\cite{Duzmanska}. 
Stanney et al.~\cite{Stanney} has found high correlations between exposure time and cybersickness, with longer exposure times increasing the risk of cybersickness.  
On the other hand, there is also the opposite view that people may build up a resistance or adapt over time (or by frequent exposures) to cybersickness~\cite{Duzmanska}.  
Thus, the exact relationship between extended exposure and cybersickness symptoms is not firmly established.

To our knowledge, no prior work on applying balance training as a way to train for tolerance to cybersickness has been reported.  
Note that similarly to any external stimulation, the mere repeated and prolonged exposure to VR in itself can certainly have the effect of insensitization or habituation to the cybersickness~\cite{palmisano2022reductions}. 
However, we expect it to be a relatively time-consuming method and quickly receding in its effect (compared to active training), and little is known about whether there is any transfer effect to other contents (for which the user was not exposed to)~\cite{palmisano2022reductions, duzmanska2018can, adhanom2022vr, smither2008reducing}.

%% file: sections/03_Experiment.tex
\section{Experiment}

\subsection{Experimental Design}
The main purpose of the experimental study is to confirm the effect whether the learned balance ability has an impact on developing one's tolerance to cybersickness.
The balance training may occur in either a non-immersive environment or a VR environment, using sickness-eliciting contents (e.g., navigation). 
We hypothesize that given the same content, the effects of the balance training on tolerance to cybersickness will be stronger if the training occurred in the VR environment compared to using the non-immersive environment (even if the given content may be different from the one used for training). 
On the other hand, to single out the effect, if any, of the balance training to cybersickness tolerance, from that of by the media type, the training method by mere extended exposure to the same sickness-eliciting VR contents must be tested too.
Humans can become habituated, desensitized, and tolerant to cybersickness after long exposure to various stimuli by VR~\cite{Fransson, Duzmanska}. 
Thus, the experiment was designed as a two-factor repeated measure between subjects; the first factor being the 
training method in three levels (as shown in Figure~\ref{fig:teaser}): 

\begin{itemize}
    \item \textbf{2DT}: watching a sickness-eliciting navigation content on a 2D projection display while carrying out a balance training routine.
    \item \textbf{VRT}: watching a sickness-eliciting navigation content using a VR headset while carrying out a balance training routine.
    \item \textbf{VRO}: only exposure/just watching a sickness-eliciting navigation content using a VR headset, but without any balance training.
\end{itemize}

\begin{figure}[t]
\centering
\subfigure[
Forest Trail for Week 1 (EW1)
]{\includegraphics[width=1\linewidth]{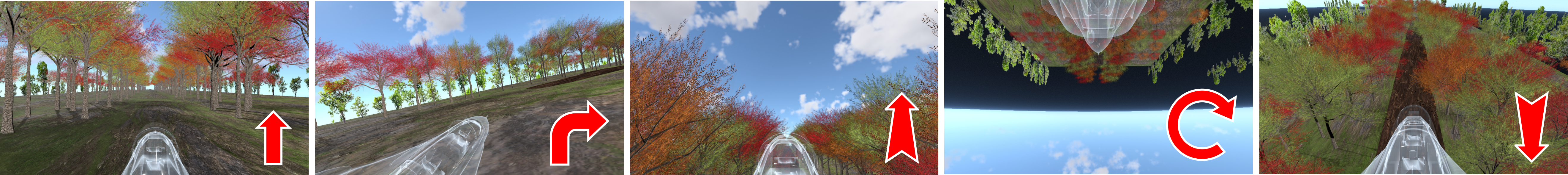}}
\hspace{3em}
\subfigure[
Wild Rollercoaster for Week 2 (EW2)
]{\includegraphics[width=1\linewidth]{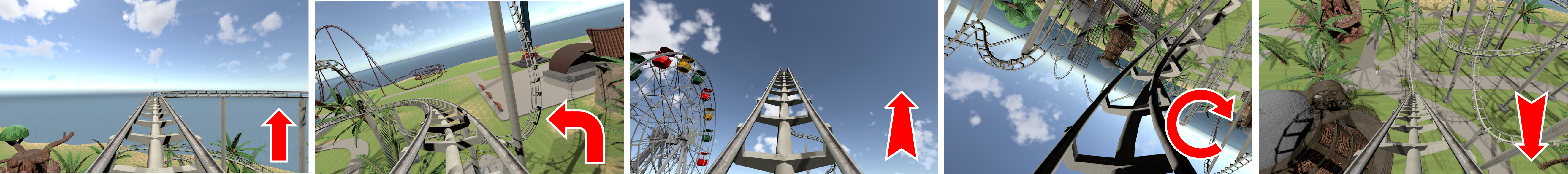}}
\caption{Snapshots depicting the different trajectories of the two balance/sickness training contents: forest trail for week 1 (EW1) and wild rollercoaster for week 2 (EW2). 
During the experiment, subjects experience the following types of movements:
    (1) straight-forward movement; 
    (2) turning left/right; 
    (3) going up;
    (4) vertical loop;
    and (5) going down.
 \label{fig:firstweek}
}
\end{figure}

\begin{figure}[ht]
\centering
\subfigure[
EW1: Relatively less sickness
]{\includegraphics[width=0.40\linewidth]{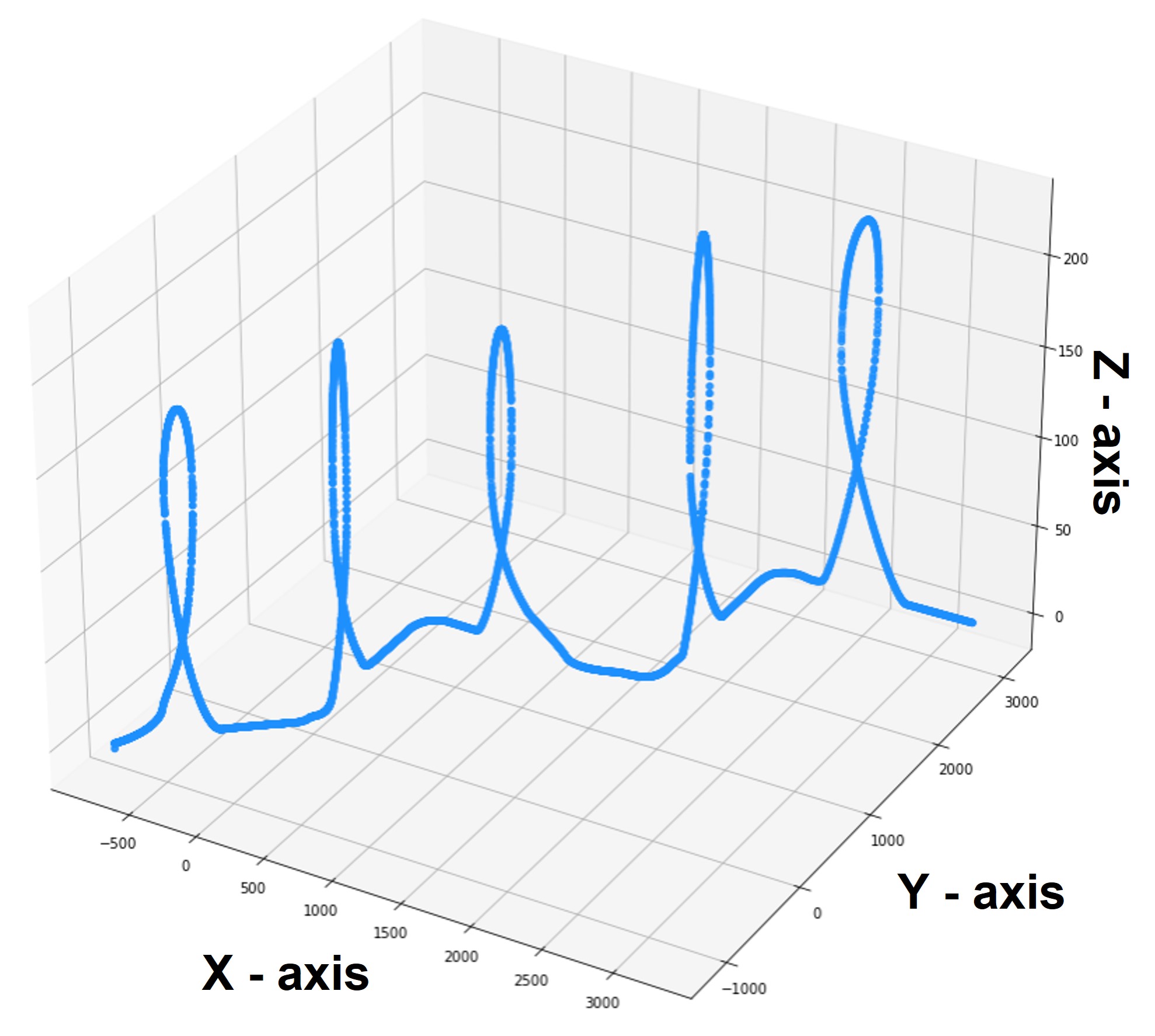}}
% \hfill
\hspace{3em}
\subfigure[
EW2: Relatively more sickness
]{\includegraphics[width=0.40\linewidth]{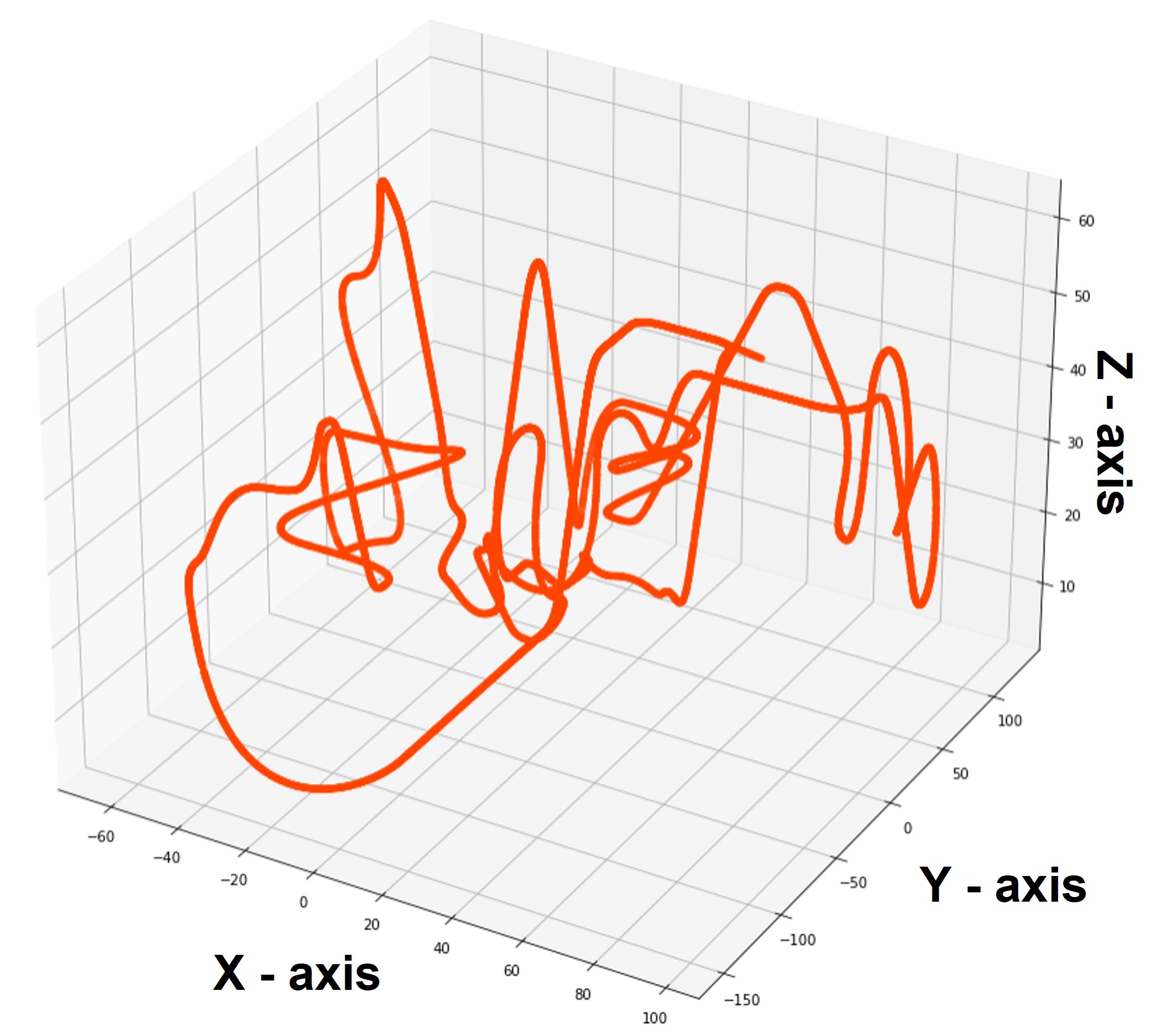}}
\caption{The navigational path profiles of the contents used in EW1 (a) and EW2 (b).
\label{trajectories}
}
% \vspace{-3em}
\end{figure}

To avoid any learning effect with regard to the contents used, a between-subject experiment was chosen.
As the effects of training may take time, the experiment was conducted over 2 weeks, but in two separate weekly segments: Experiment Week 1 (EW1) and 2 (EW2).  
Note that the same subject groups of EW1 continued to participate in EW2.  
Thus, the time (days) constituted the second factor.  
Two weeks of balance training was deemed sufficient because marked progress is usually attainable in that time frame~\cite{RASOOL2007177, SZCZERBIK2021513}. 
EW1 proceeded over 4 days, and the subjects were trained while watching the VR content which induced only a relatively moderate/less degree of sickness as to start the overall training gently (not too abruptly).  
% \redout{Subjects were made sure and assumed to have had no prior relevant training.} 
After a three-day break, EW2 was conducted with a duration of 5 days.
Due to the possible learning effect and getting accustomed to the same content after repeated exposures, a new and more dynamic content with a relatively higher degree of sickness was used (see Figure~\ref{fig:firstweek}).  
Although it is difficult to exactly quantify the difference in the sickness levels, from Figure~\ref{trajectories} which shows the navigation motion profiles of the respective content, it is reasonably clear that the Week 2 content would induce a much more severe level of cybersickness. 
EW2 was administered three days after the 4-day EW1 with the treatment-wise same subjects from EW1, thus we postulate that that subjects still were possibly affected by the training given in EW1. 

% \begin{figure*}[h]
%     \centering
%     \includegraphics[width=\linewidth]{figures/1week-track-images.jpg}
%     \vspace{-1em}
%     \caption{
%     \label{fig:firstweek}}
%     \vspace{-1em}
% \end{figure*}

In summary, there were two mixed model and longitudinal experiments; each designed as two factors, 3 x 2, repeated measure between subjects. Even though experimental tasks were carried out and dependent variables measured every day during the 4-day/5-day periods for EW1/EW2, we focus only on the difference between the first and last days and data analyzed accordingly (making it a two-level study for the second factor).
Figure~\ref{fig:process} shows the overall experimental process.
The hypotheses regarding the outcome of the experiment can be summarized as follows:

\begin{figure}
    % \centering
    \includegraphics[width=0.7\columnwidth]{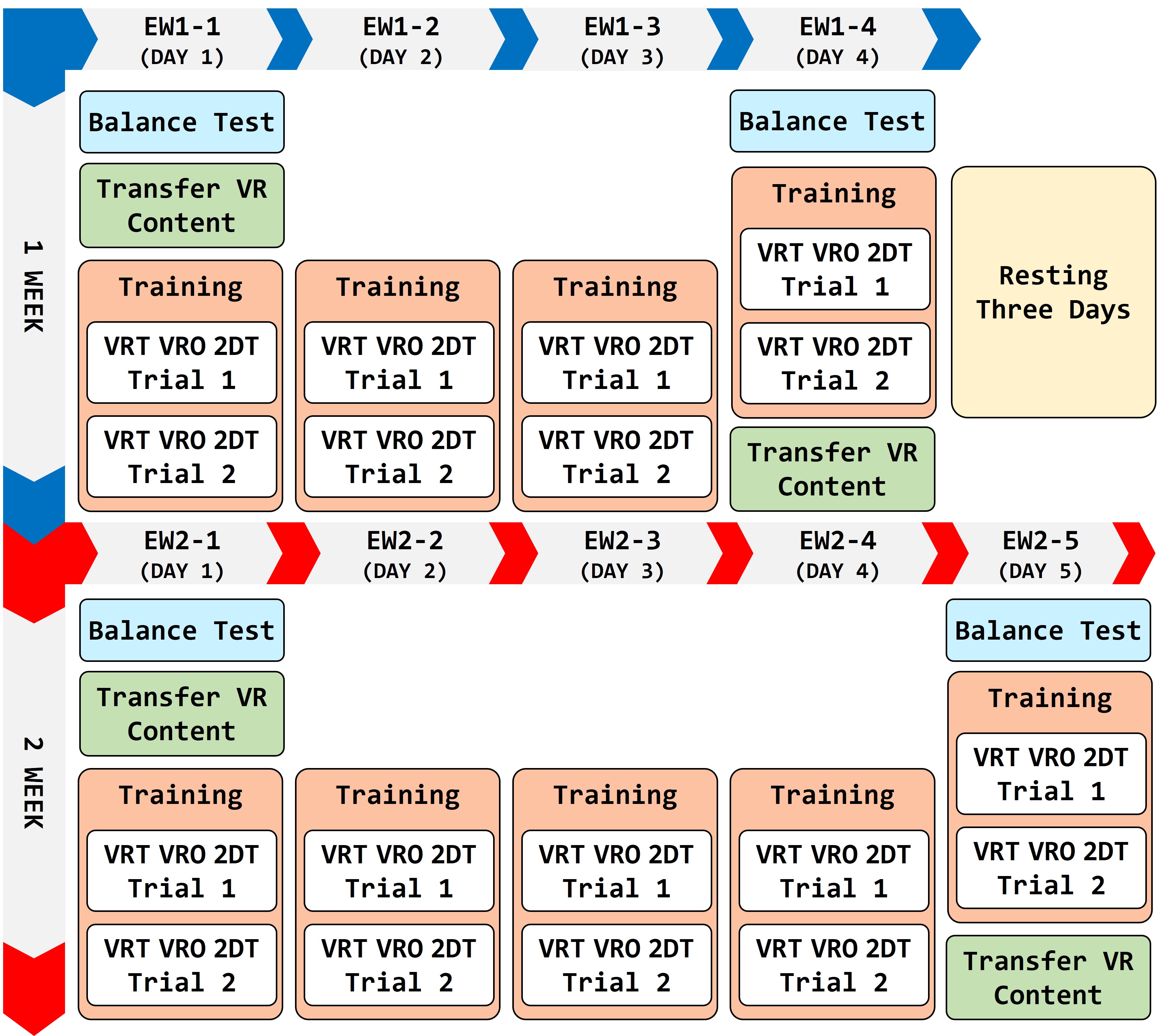}
    \caption{The overall experimental process for EW1 and EW2.~\label{fig:process}}
\end{figure}

\begin{itemize}
    \item \textbf{H1}: There will be a balance training effect
    (i.e., significantly getting better in time) in both immersive VR (VRT) and 2D-based environments (2DT). 
    
    \item \textbf{H2}: Balance training improves tolerance to cybersickness. If so, this partly serves as evidence for the posture instability theory where imbalance can be considered the cause of cybersickness.
    
    \item \textbf{H3}: The training effect for balancing and cybersickness will be greater with the use of immersive VR (VRT) than with the 2D environment (2DT). 
    
    \item \textbf{H4}: The training effect for cybersickness through balance training (with VR and/or non-VR) will be greater than just the extended exposure to the same content.
\end{itemize}

\subsection{Experimental Task and Training Contents}
In both EW1 and EW2, except for the training contents used, 
the experimental task and procedure were the same.
For 2DT and VRT, 
part of the experimental task was to follow a simple balance training routine called the 
``one leg stand'' (also known as the Flamingo test~\cite{Uzunkulaoglu, Marcori}).
In 2DT, the balance training was administered while watching the navigational content on the projection display (60 inches viewed from 1.5 m away), 
and in VRT, watching the same content through the VR headset (Oculus Quest 2 VR headset with a FOV of 104 $x$ 98 degrees).
During the 3-minute experience, the balance training routine is as follows: ready/rest (30s') - training (30s') - rest (30s') - training (30s') - rest (30s') - training (30s'). 
The instructions are presented through a distinct and visible user interface in the VR/2D projection system.
% The 3-minute balance training routine consists of alternating periods of 30 seconds each for Ready/Rest, training, and rest, with instructions provided through a distinct visible UI in the VR/2D-projection system.
% During the 3-minute experience, the balance training routine is as follows: Ready/Rest (30s') - training (30s') - Rest (30s') - training (30s') - Rest (30s') - training (30s').
The instructions were provided by the distinct visible UI of the VR/2D-projection system.
For VRO, no balance training occurred - the subject just watched the same VR content with the VR headset standing on two feet.

As already indicated, 
two contents of different sickness levels and themes were used 
- the lesser sickness eliciting one in EW1 (flying through the forest trail) 
and more in EW2 (wild roller coaster ride).  
Figure~\ref{fig:teaser_contents} (a), (b), and Figure~\ref{fig:firstweek} show the example scenes from the respective contents. 
Experimenting with the new more difficult (sickening) content also provided the opportunity to examine the user behavior and performance after a week of training.  

% For both contents, the viewpoint was set slightly behind the agent representing the user 
% (i.e., a point of view that is neither fully first-person nor third-person)
% % (1.5 persons's point of view).
% so that it was visible without having to rotate one's head too much to see oneself which we regard to be important visual feedback in the balance training.
The navigation path contained several types of motion - forward translation and pitch/yaw, rotation/turning in varied speed and acceleration, as shown in Figure~\ref{fig:firstweek}, - mixed up for each content to be clearly differentiated in their respective sickness level (also informally tested with no subjects).
In addition to using different contents between EW1 and EW2 to prevent the learning effect, 
similar provisions were made within the same content within EW1 and EW2 as well.  
Each content was presented in slightly different versions by varying the navigation path/motion profiles and surrounding environment objects (while maintaining the same sickness level) between each day.
% \textcolor{red}{
% In addition, in the navigation content obstacles appeared, from time to time, with openings for which the subject had to take a particular pose (e.g. T pose) to successfully pass through without penalty (all while trying to maintain balance in one leg).  ... strike out ...  
% }

The training proceeded 2 times a day for four days in EW1 and likewise for five days in EW2.
The subject was free to put one's foot back down anytime if felt to be in danger of falling down (or for any reason e.g., whether not able to maintain balance or due to too much sickness) but was asked to resume and continue in one's best way. 
The experiment helper stood by to prevent the subject from completely falling down. 
The subject was also free to stop the experiment at any time, although there was no such case.

\vspace{-0.5em}
\subsection{Dependent Variables}

The dependent variables of main interest were two: changes in the balance performance and cybersickness scores over time.  
Each of these aspects was measured in several ways.

The balancing performance was measured quantitatively by 
(1) the number of times the subject put back one's foot down during the training process, 
and (2) computing the extent of the deviation of the body from the reference center of mass (e.g., when standing still).  
The former was manually counted, while the latter, was obtained off-line by analyzing the subject's 2D pose data extracted from the recorded video using the PoseNet \cite{Papandreou_2017_CVPR,Papandreou}. 
In particular, the variation of the midpoint of the screen space locations of the right and left hips were used to estimate this measure. 
We omit further implementation details for lack of space.  

In addition, 
to assess whether the balancing performance improved regardless of the given training content, 
we measured the subject's performances of the ``one leg stand with eyes closed''~\cite{Bohannon} on the first and last days of each week (see Figure~\ref{fig:process}).

As for the overall cybersickness level assessment, Kennedy's Simulation Sickness Questionnaire (SSQ) score was used~\cite{kennedy1993simulator}.
However, since the SSQ only asks for the existence of certain symptoms, it is not possible to assess their probable cause for our experiment - e.g., from the visual motion or balancing act. 
Thus, in addition to the original SSQ (herein referred to as the ``Original''), two revised versions, called the ``Visual'' and ``Balance'', were made and used.
Questions in the revised versions ask of the same symptoms, but also of what the subjects thought the source might be, i.e., from the visual motion or the one-leg stand balancing act. 
The whole questionnaire comprised 48 questions (see Appendix).

\begin{figure*}[h]
\centering
\subfigure[Rollercoaster ride]{\includegraphics[trim={0 0 0 0}, clip, width=0.49\linewidth]{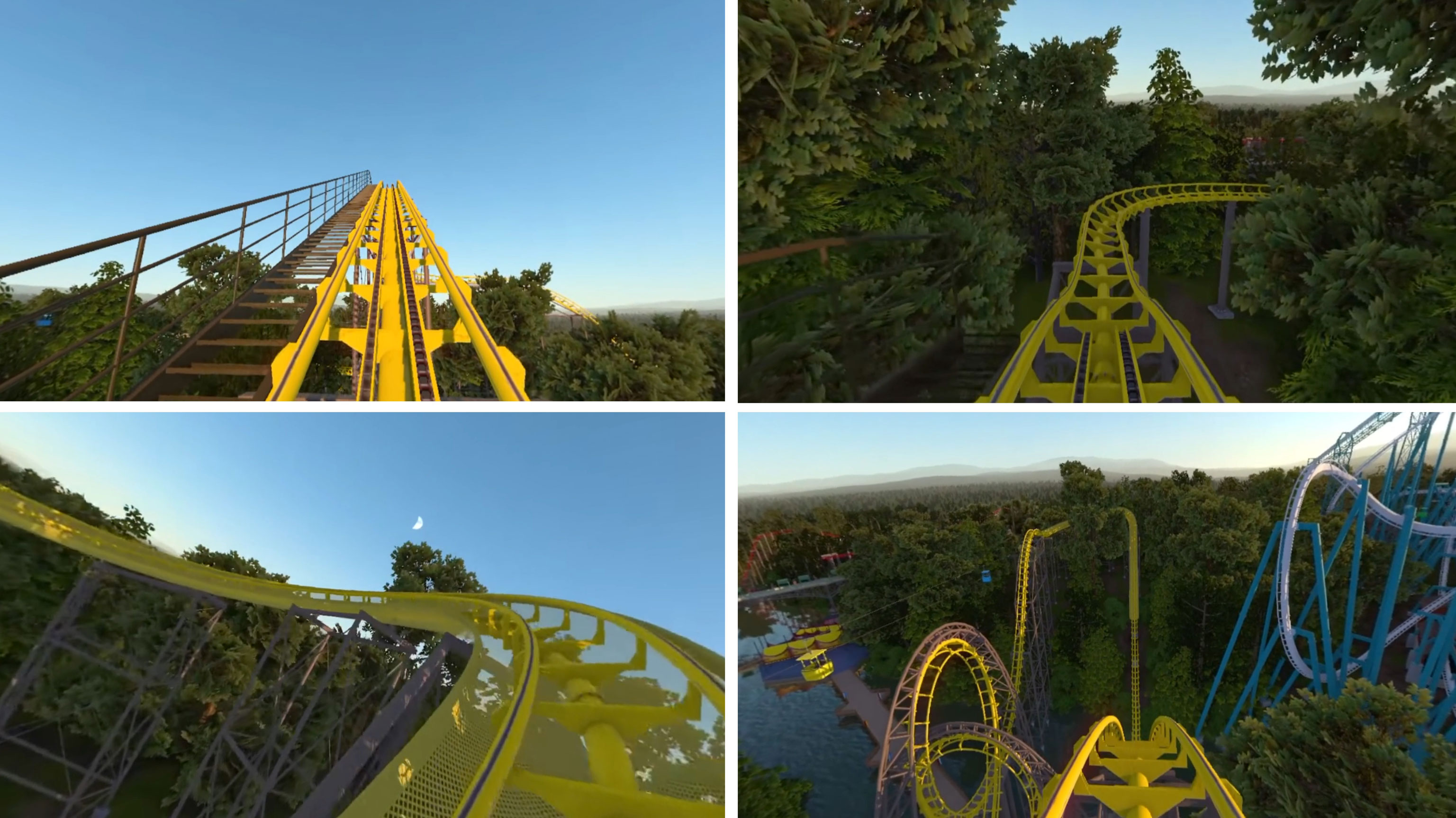}}
\hfill
\subfigure[Space navigation]{\includegraphics[trim={0 0 0 0}, clip, width=0.49\linewidth]{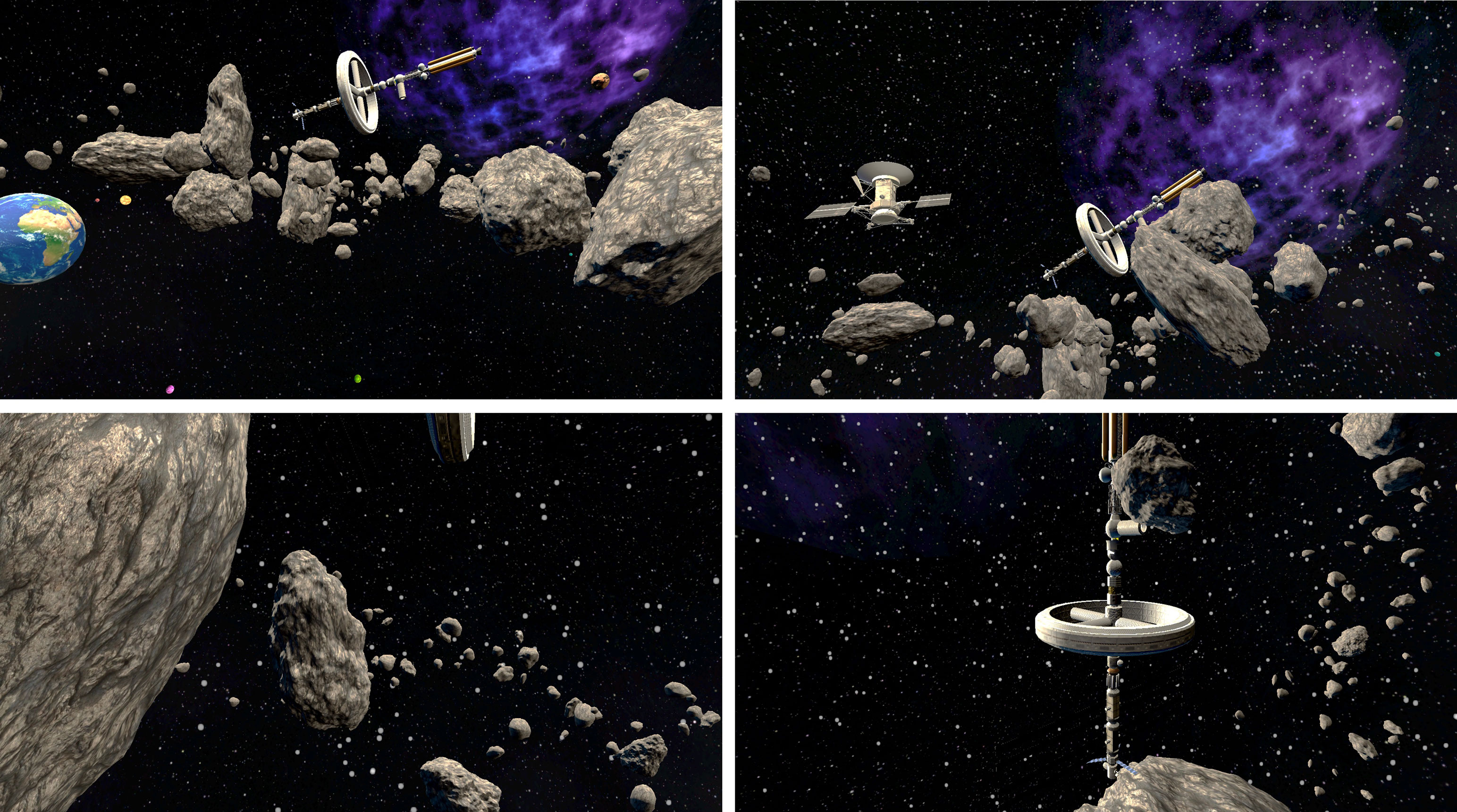}}
\caption{
The two contents to test the transfer effect of balance training: (a) rollercoaster ride before and after EW1 and (b) space exploration before and after EW2.\label{transfereffectvirtualenv}
}
% \vspace{-0.5em}
\end{figure*}

Lastly, to assess whether the cybersickness tolerance improved regardless of the given content (i.e., transfer effect to another VR content), 
we measured the subject's sickness levels using a completely different (from the ones used in the main experiment) sickness-inducing content; 
tested with transfer VR content 1 (rollercoaster ride\footnote{
The YouTube 360-degree video link is available at \url{https://www.youtube.com/watch?v=eHAu8BV85vE}.}, and transfer VR content 2 (space exploration) before and after EW2.
\note{The duration of both content was 3 minutes, the same as the experimental content.
Moreover, subjects in all conditions experienced the transfer content while standing and wearing a VR headset without any balance training.}
We reemphasize that the rollercoaster content used to explore the transfer effect was completely different from the one used for training in EW2.
These transfer effect test contents are illustrated in Figure~\ref{transfereffectvirtualenv}. 
% \note{Figure 4 illustrates the entire experimental task and measurement procedure.  }

\begin{table}[t]
\renewcommand{\arraystretch}{1.1}
\caption{Dependent variables}
\centering
\begin{tabular}{ll}
\toprule
\textbf{Category}            & \textbf{Dependent Variable}        \\ \midrule
\textbf{Balance Performance}          & Number of times foot was put down \\
                             & Time maintaining one leg stand \\
                             & Center of mass variability \\ \midrule
\textbf{Cybersickness}                  & Original SSQ               \\
                             & Visual SSQ \\
                             & Balance SSQ \\ \midrule
\textbf{Transfer Effect}              & Original SSQ \\ 
\bottomrule
\end{tabular}
\end{table}

%Figure 1 Daily Routine으로 좌측 도표는 VRT, 2DT에서의 루틴, 우측 도표는 VRO의 루틴을 나태낸다. 1일, 5일, 6일차에 사전 세션이 루틴에 포함되며, 10일차에 사후 세션이 진행되었다. 사전, 사후 세션은 트레이닝에서 사용된 콘텐츠와 다른 콘텐츠로 멀미감을 측정하였다

% \begin{table}[h]
% \centering
% \vspace{-0.5em}
% \caption{The experimental task and measurement procedure over EW1 and EW2. 
% \label{table:expdes}}
% \vspace{-0.5em}
% \begin{tabular}{ccccc}
% \toprule
% Day 1 & Day 2   & Day 5 & Day 6    & Day 10 \\ 
%       & - Day 4 &       & - Day 9  &        \\ \midrule
% Blind 1 leg & Forest trail & Blind 1 leg & Wild RC & Blind 1 leg \\
% VR sick &             & VR sick  &                 & VR sick \\
% test 1 &             & test 1 and 2 &                 & test 2 \\
% \bottomrule
% \end{tabular}
% \vspace{-1em}
% \end{table}

\subsection{Participants}

Subjects were initially recruited through the university's online community. The first round of subjects was surveyed for their self-reported sensitivity to motion sickness using the MSSQ-short~\cite{golding2006predicting, NESBITT20171} and familiarity or prior experiences in using the VR system.  
We notified the potential subjects of the need to carry out balance training (one leg stand) for about 10-15 minutes per day for two weeks and asked them to excuse themselves if they deemed it to be beyond their physical capabilities.
Subjects in the extreme ends in terms of their reported sensitivity were also excluded, as our study targeted subjects in the middle of the sensitivity spectrum. 

Fifteen final subjects were selected and placed in the three subject groups (5 each) for 2DT, VRO, and VRT such that their MSSQ score variations were similar and within an acceptable range (all male, aged 19 to 33, mean = 25.6, SD = 2.19). 
All participants had at least some experience in using VR applications (mostly game playing and video watching) but did not have any prior balance training experience.
The subjects were paid 16 USD per hour for their participation (a total of about \$ 120 for the whole two weeks). 
All 15 subjects managed to finish the experiment in two weeks without giving up in the middle.

% mean height of 172.9 $cm$ (SD = 4.2) 
% and an average weight of 74.7 $kg$ (SD = 14.87)

%피험자의 모집은 대학교의 온라인 커뮤니티를 통해 30명 이상의 실험 참가 후보자들을 모집하였다. 모집된 실험 대상자들은 사전 자가진단 설문조사인 MSSQ를 작성하여 평소에 느끼는 멀미감과 가상현실에서의 멀미감의 민감도, 그리고 VR에 대한 친숙도를 평가받았다. 그 후 VRT, VRO, 2DT의 각 실험군에 MSSQ 스코어의 표준편차 이상 차이나지 않도록 균형을 맞추었다. 또한 한 다리를 들고 버티는 트레이닝 자세를 지속적으로 수행하야 하므로 신체 운동에 무리가 있는 피험자와 전정기관에 장애가 있는 피험자는 제외하였다. 

% 2주간의 연속적인 트레이닝 실험으로 인해 피험자를 15명으로 제한하였고 VRT, VRO, 2DT 실험 집단에 각 5명씩 배정되었으며(남성 15명, 평균 나이 25.6, 표준편차 2.19)  모든 피험자는 VR 사용 경험이 있었다.

%  실험 참여자들은 실험 비용을 하루 당 15,000원으로 열흘 모두 참여 시 150,000을 제공받았다. 중도 포기자는 없었으며, 10일간의 스케줄에 피험자 모두 참여하였다. 실험 시작 전 피험자들은 연구참여 동의서를 작성하였고, 실험에 대해 충분히 인지할 수 있도록 상세한 설명이 제공되었다. 특히 Overall, Visual, Balance로 구분된 SSQ 설문 문항에 대해 각 문항이 서로 배타적이지 않음을 인지하도록 하였다. 

%낙상의 위험을 방지하기 위해서 바닥에 유아용 메트리스를 설치하였으며, 양 옆에 메트리스를 세워 혹시 모를 낙상에 대비하였다. 피험자에게 실험 전 낙상할 것 같은 경우 두 발을 땅에 지지하도록 요청하여 실제 실험에서는 단 한번의 낙상도 발생하지 않았다. 또한 안전요원이 항시 앞에서 대기하여 피험자가 충분히 안전한 상황에서 실험에 참여할 수 있도록 하였다.

\subsection{Experimental Set-up and Procedure}

The subjects first filled out the consent agreement form, were briefed about the \note{procedure of the experiment}, and explained the experimental tasks. 
Five to ten minutes were given for the subjects to get oneself familiarized with the balancing task  
while watching the content through the monitor or the headset. 
In particular, the subjects were given detailed instructions on how to carefully respond to the three types of SSQs and to deeply think about the probable causes of the symptoms the best they could. The helper assisted the subject to position oneself in front of the monitor on the floor (with cushioned walls) or donning and adjusting the headset.  The helper also stood by to prevent the subject from falling down.

% On each day, the subjects for 2DT and VRT carried out the balance training procedure for 30 seconds three times with 30 seconds of rest in between. 

On each day, the subjects for 2DT and VRT alternated between a 30-second rest/preparation period, followed by a 30-second balance training procedure, 
repeating this sequence three times in total.
The subjects selected which foot to use to stand or lift on their own.
This protocol was designed considering that the similar Y-balance (one leg with eyes closed) test typically lasted around 30 seconds on average~\cite{ybalance}, and our own pilot test (with four males) indicated that exceeding one minute often led to muscle strain.  Although the experiments were run over two weeks and sufficient rests were taken between the treatments, minimizing subject fatigue and
ensuring the best physical condition was deemed important to derive the best and credible results.

Subjects of VRO simply watched the navigation content one time for 3 minutes in a normal standing pose. 
After the respective treatments, subjects rested and filled out the sickness survey.  
Subjects were free to stop the experiment for any reason without any financial penalty. After experiencing all treatments, informal post-briefings were taken. 
The experiment was approved by the Institutional Review Board (IRB No. 2023-0143-01).

The experimental contents were developed using the Unity 3D game engine, specifically version 2021.3.12f, and run and displayed with the Oculus Quest 2 VR headset.

%% file: sections/04_Results.tex
\section{Results}
\label{sec:results}
Considering the 3x2 mixed design of the experiments and the collected longitudinal data being both continuous and non-parametric, we analyzed the data using the nparLD~\cite{JSSv050i12} to examine the effects. 
For pairwise comparisons, we employed the Kruskal-Wallis test to analyze the factor of the training method (as a between-design factor), while the factor of the time (as a within-design factor) was assessed using the Wilcoxon signed-rank test.
All tests were applied Bonferroni correction with a 5\% significant level.

Moreover, since EW1 and EW2 were conducted with different settings and subject conditions, 
they were treated as two separate experiments for analysis.
As the same group of subjects was used treatment-wise, treatment-wise statistical and subjective comparison between EW1 and EW2 is possible (with respect to the factor of the time).  
On the other hand, the effect analysis within EW1 and EW2 with respect to the training method 
(i.e., among 2DT, VRT, and VRO) would be less reliable as the subject groups were different, and their numbers were low (only 5 in each group). 

\subsection{Change in Sickness Levels}
\label{sec:change in sickness levels}
The primary focus of our study was to alleviate cybersickness caused by visual mismatch through balance training. However, we considered the possibility that physical discomfort or hardship from the balancing act
could be similar to the many symptoms of the cybersickness as assessed by the SSQ (e.g. ``disorientation" from trying to stand on one foot).    
Therefore, we administered two additional revised versions of the ``Original'' SSQ: namely, ``Visual'', and ``Balance'' - for responders to distinguish between and report whether the cybersickness-like symptoms were induced by the visual mismatch and/or by the balance training.
Nevertheless, we do acknowledge the subjects' inherent difficulty (despite the survey's kind explanation and clarification) to objectively and correctly judging the sources of the symptoms, and also of the possible interaction between these two mixed factors.  Note that the subject was free to attribute a given symptom to both visual stimulation and balancing exercise. 
On the other hand, it is not common to see people seriously suffer from sickness-like symptoms from just doing balance exercises.  
Considering all these, our analysis focused on investigating the 
effects on the Visual SSQ scores.

\begin{figure}[t]
\centering
\subfigure[EW1: Week 1]{\includegraphics[trim={0 0 0 0}, clip, width=0.48\columnwidth]{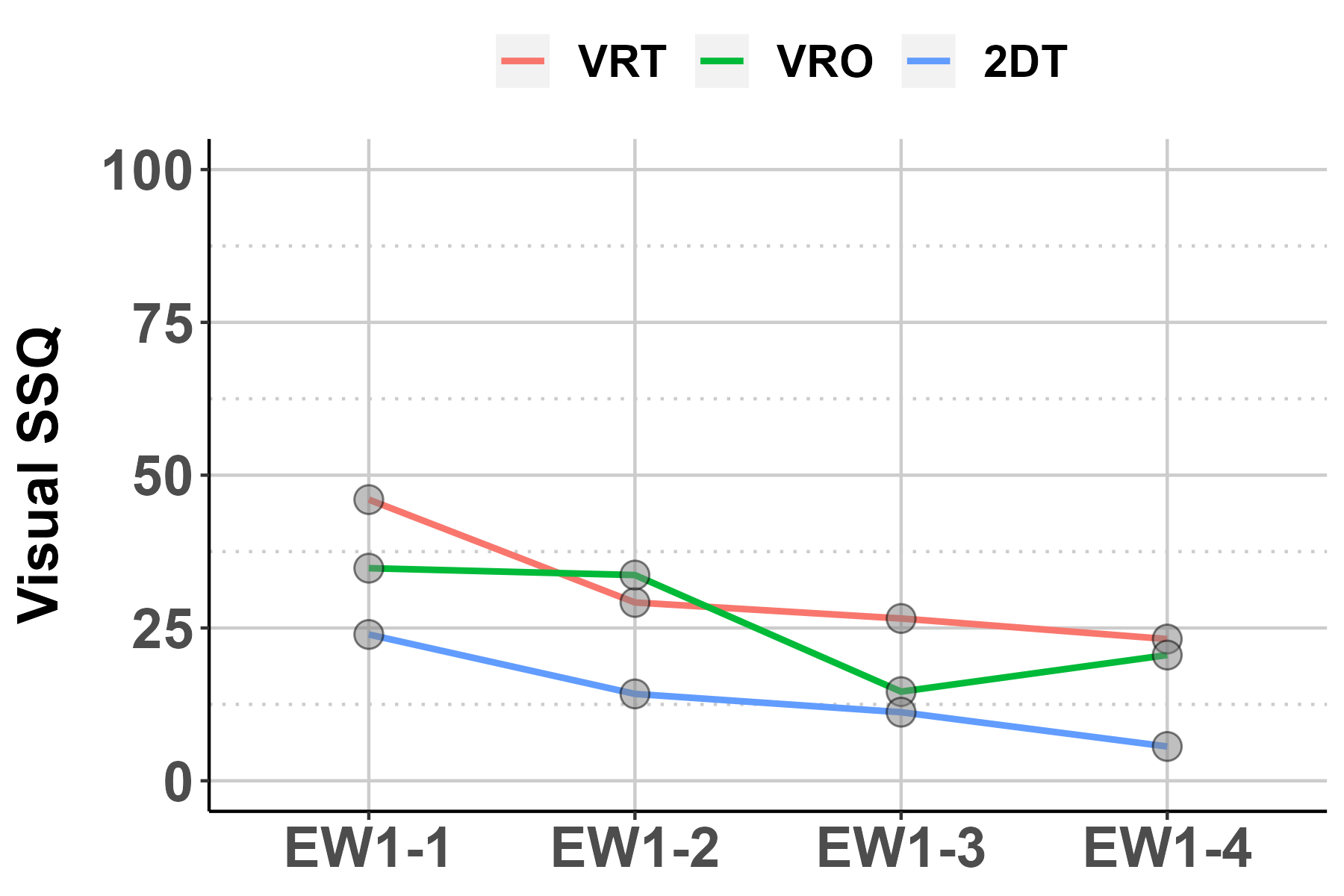}}
\subfigure[EW2: Week 2]{\includegraphics[trim={0 0 0 0}, clip, width=0.48\columnwidth]{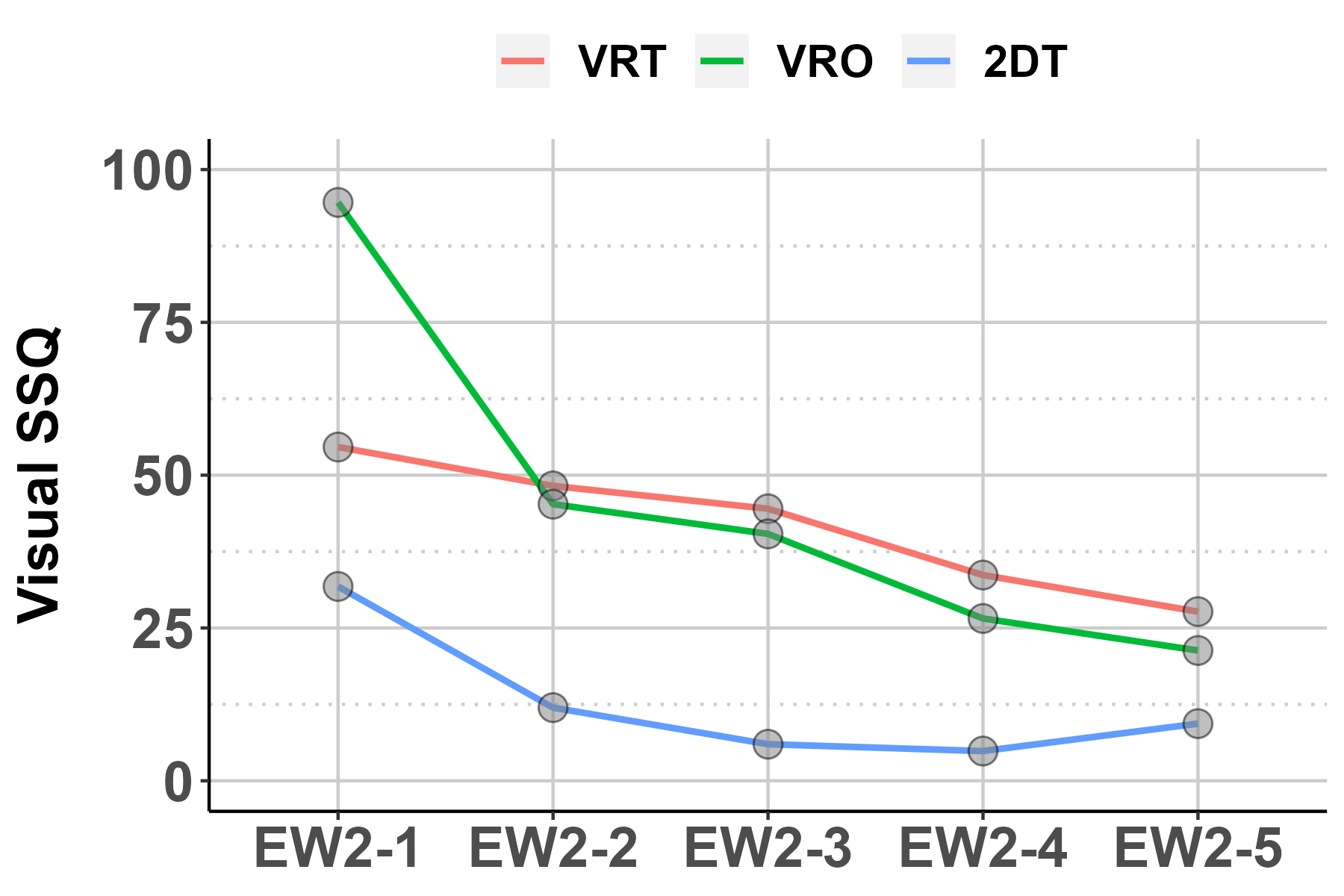}}
\caption{
Changes in the average Visual SSQ's total scores over the 4-day and 5-day periods of EW1 and EW2.
~\label{AverageScroeOfSSQ}
}
\end{figure}

Figure~\ref{AverageScroeOfSSQ} (and also partly for just the first and last days in Table~\ref{SSQ overall})
illustrates the trends of the average Visual SSQ scores over the 4-day/5-day periods of EW1 and EW2 for VRT, VRO, and 2DT.  
The detailed statistical figures, including the pairwise comparisons, are given in Figure~\ref{nparLDananalysisResutls}, Table~\ref{table:nparLD}, Table~\ref{kruskal}, and Table~\ref{wilcoxon}.

\begin{table}[t]
\renewcommand{\arraystretch}{1.5}
\caption{Average total SSQ scores over different training methods and the p-value from the Wilcoxon sign rank test on the first and last days of EW1 and EW2
(* $p$\,\textless\,.05; ** $p$\,\textless\,.01; and *** $p$\,\textless\,.001). \label{SSQ overall}}
\resizebox{\linewidth}{!}{
\centering
\begin{tabular}{llrrlrrl}
\hline \hline
                     &  & \multicolumn{1}{l}{\textbf{EW1-1}} & \multicolumn{1}{l}{\textbf{EW1-4}} &  \textit{p-value} &  \multicolumn{1}{l}{\textbf{EW2-1}} & \multicolumn{1}{l}{\textbf{EW2-5}} & \textit{p-value}\\ \hline
\textbf{Original SSQ} & \textbf{VRT}       & 54.604         & 24.31     &   $<$ .001 ***  & 51.612       & 27.676     &    .004 **                      \\
                     & \textbf{VRO}       & 35.904         & 21.318    &   .092  & 94.996             & 21.692     &   $<$ .001 ***                    \\
                     & \textbf{2DT}       & 35.156         & 13.464    &   .007 **  & 32.164           & 13.838     &   .003 **                     \\ \midrule
\textbf{Visual SSQ}  & \textbf{VRT}       & 46.002         & 23.188    &   .021 *  & 54.604            & 27.676     &   .004 **                      \\
                     & \textbf{VRO}       & 34.782         & 20.57     &   .092  & 94.622             & 21.318     &   $<$ .001 ***                     \\
                     & \textbf{2DT}       & 23.936         & 5.61      &   .025 *  & 31.79             & 9.35       &   $<$ .001 ***                     \\ \midrule
\textbf{Balance SSQ} & \textbf{VRT}       & 36.652         & 18.7      &   .020 *  & 36.278            & 20.944     &   .092                    \\
                     & \textbf{VRO}       & 7.106          & 13.09     &   .70  & 43.01               & 1.496      &   .071                     \\
                     & \textbf{2DT}       & 40.392         & 8.602     &   $<$ .001 *** & 16.456        & 8.228      &   .054                   \\ \hline \hline
\end{tabular}
}
\end{table}

\begin{table}[t]
\renewcommand{\arraystretch}{1.25}
\caption{Statistical analysis results for the effects on Visual SSQ for the training content and 
transfer content in EW1 and EW2, respectively (* $p$\,\textless\,.05; ** $p$\,\textless\,.01; and *** $p$\,\textless\,.001).~\label{table:nparLD}}
\centering
\begin{tabular}{lllll}
\hline \hline
\textbf{EW}    & \textbf{Variable}     & \textbf{Factor}                & \textit{F}     & \textit{p}         \\ \hline
EW1   & Visual SSQ   & Training Methods      & 3.16  & .045 *             \\
      &              & Days                  & 11.23 & \textless .001 *** \\
      &              & Training Methods:Days & .02   & .97                \\ 
      & Transfer SSQ & Training Methods      & 1.44  & .23                \\ 
      &              & Days                  & .97   & .32                \\
      &              & Training Methods:Days & .25   & .71                \\ \hline
EW2 & Visual SSQ   & Training Methods      & 3.66  & .029 *             \\
      &              & Days                  & 90.63 & \textless .001 *** \\
      &              & Training Methods:Days & 3.7   & .025 *             \\ 
      & Transfer SSQ & Training Methods      & 1.58  & .20                \\ 
      &              & Days                  & 8.05  & .004 **            \\
      &              & Training Methods:Days & 0.41  & .64                \\ \hline \hline
\end{tabular}
\end{table}

\subsubsection{EW1}

In EW1, according to the nparLD~\cite{JSSv050i12}, 
we observe significant differences in the level of cybersickness (Visual SSQ) in relation to both the training method ($p <$ .05) and the days ($p <$ .001), but no interaction effect between the training method and the days (see Table~\ref{table:nparLD}).
However, the pairwise comparison in Table~\ref{kruskal}
shows no statistically significant differences
on the factor of the training method on the first or last days of the experiments.
These results might be attributed to the training method being a between-subject factor, 
with its impact being relatively small and less reliable (plus the difference in the nparlD and Kruskal analyses).

\begin{table}[t]
\renewcommand{\arraystretch}{1.25}
\caption{
Pairwise comparison: Between-subject factors (VRT vs. VRO vs. 2DT) using the Kruskall-Wallis test. 
The lower group had less sickness (* $p$\,\textless\,.05).~\label{kruskal}}
\centering
\begin{tabular}{llllll}
\hline \hline
\textbf{EW}    & \textbf{Variable} & \textbf{Days}  & \textbf{$\chi^2$}  & \textit{p}    &                 \\ \hline
EW1 & Visual SSQ         & EW1-1              & 3.71 & .156 &                         \\
      &                    & EW1-4              & 3.43 & .180 &                         \\
      & Transfer SSQ       & EW1-1              & 2.92 & .233 &                         \\
      &                    & EW1-4              & 1.83 & .40  &                         \\ \hline
EW2 & Visual SSQ         & \textbf{EW2-1}              & \textbf{9.11} & \textbf{.011} & \textbf{VRO \textgreater 2DT *} \\
      &                    & EW2-5              & 3.61 & .164 &                         \\
      & Transfer SSQ       & EW2-1              & .92  & .63  &                         \\
      &                    & EW2-5              & 4.12 & .128 &                         \\ 
\hline \hline
\end{tabular}
\end{table}

\begin{table}[t]
\renewcommand{\arraystretch}{1.2}
\caption{Pairwise comparison: Within-subject factors (First day vs. Last day) 
using the Wilcoxon-sign rank test. 
The lower group had less sickness (* $p$\,\textless\,.05; ** $p$\,\textless\,.01; and *** $p$\,\textless\,.001).~\label{wilcoxon}}
% \vspace{-0.5em}
% \resizebox{\linewidth}{!}{
\centering
\begin{tabular}{llllll}
\hline \hline
\textbf{EW}    & \textbf{Variable} & \textbf{Methods}  & \textit{Z}  & \textit{p}    &                 \\ \hline
EW1 & Visual SSQ         & VRT              & -2.04   & .02             &  1 day $>$ 4 day *                        \\
      &                    & VRO              & -1.32   & .09             &                         \\
      &                    & 2DT              & -1.9    & .02             &  1 day $>$ 4 day *                      \\
      & Transfer SSQ       & VRT              & -0.40   & .42             &                         \\
      &                    & VRO              & 0.94    & .20             &                         \\ 
      &                    & 2DT              & -0.13   & .29             &                         \\ \hline
EW2 & Visual SSQ         & VRT              & -2.65   & .004            & 1 day $>$ 5 day **                        \\
      &                    & VRO              & -3.09   & \textless .001  & 1 day $>$ 5 day ***                       \\
      &                    & 2DT              & -3.09   & \textless .001  & 1 day $>$ 5 day ***                        \\
      & Transfer SSQ       & VRT              & 2.02    & .02             & 1 day $>$ 5 day *                        \\
      &                    & VRO              & 0.94    & .20             &                         \\ 
      &                    & 2DT              & 0.40    & .1              &                         \\ 
\hline \hline
\end{tabular}
% }
\end{table}

\begin{figure*}[h]
\centering
\subfigure[EW1 Visual SSQ]{\includegraphics[trim={0 2em 3em 0em}, clip, width=0.24\linewidth]{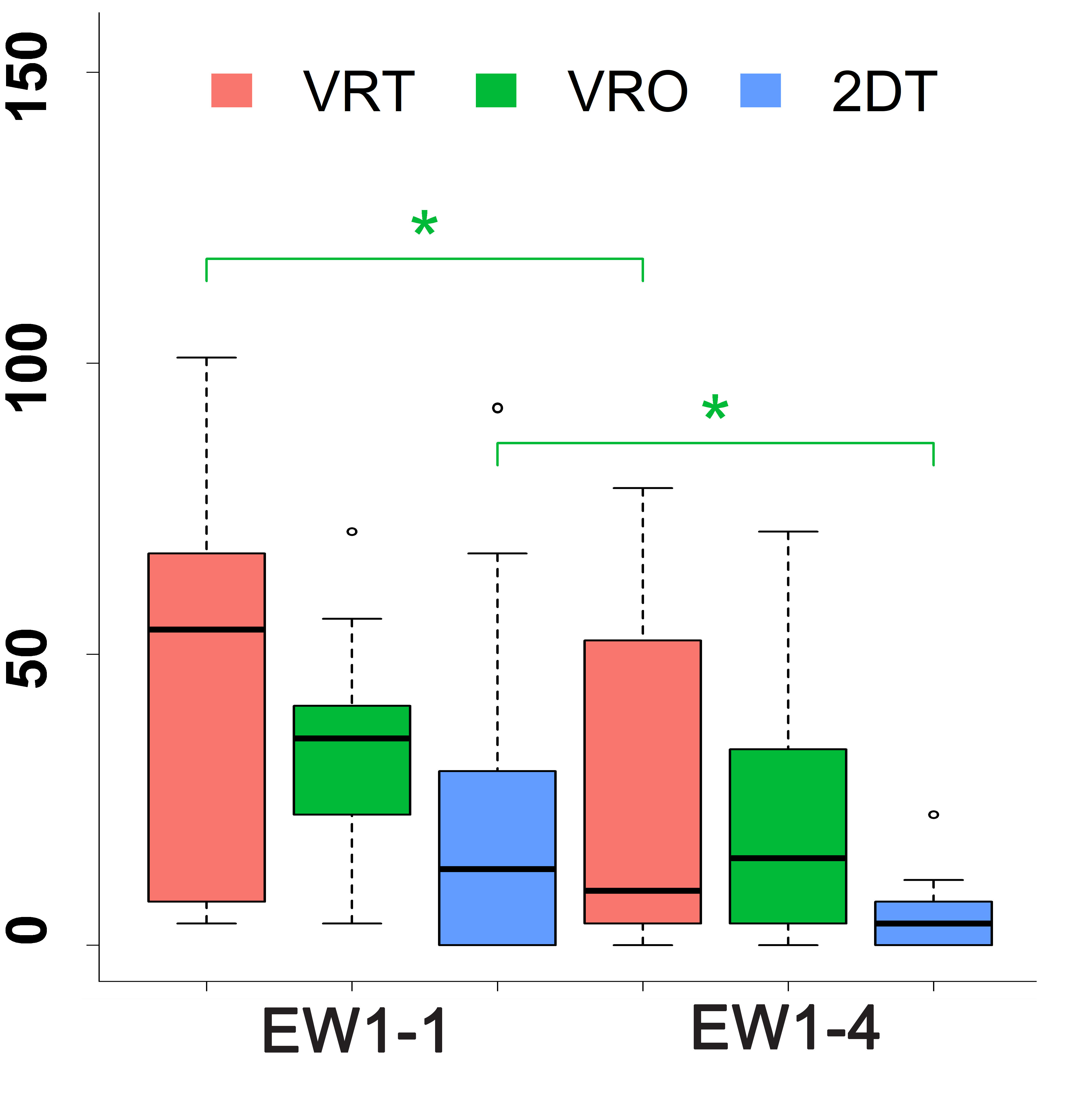}}
\hfill
\subfigure[EW2 Visual SSQ]{\includegraphics[trim={0 2em 3em 0em}, clip, width=0.24\linewidth]{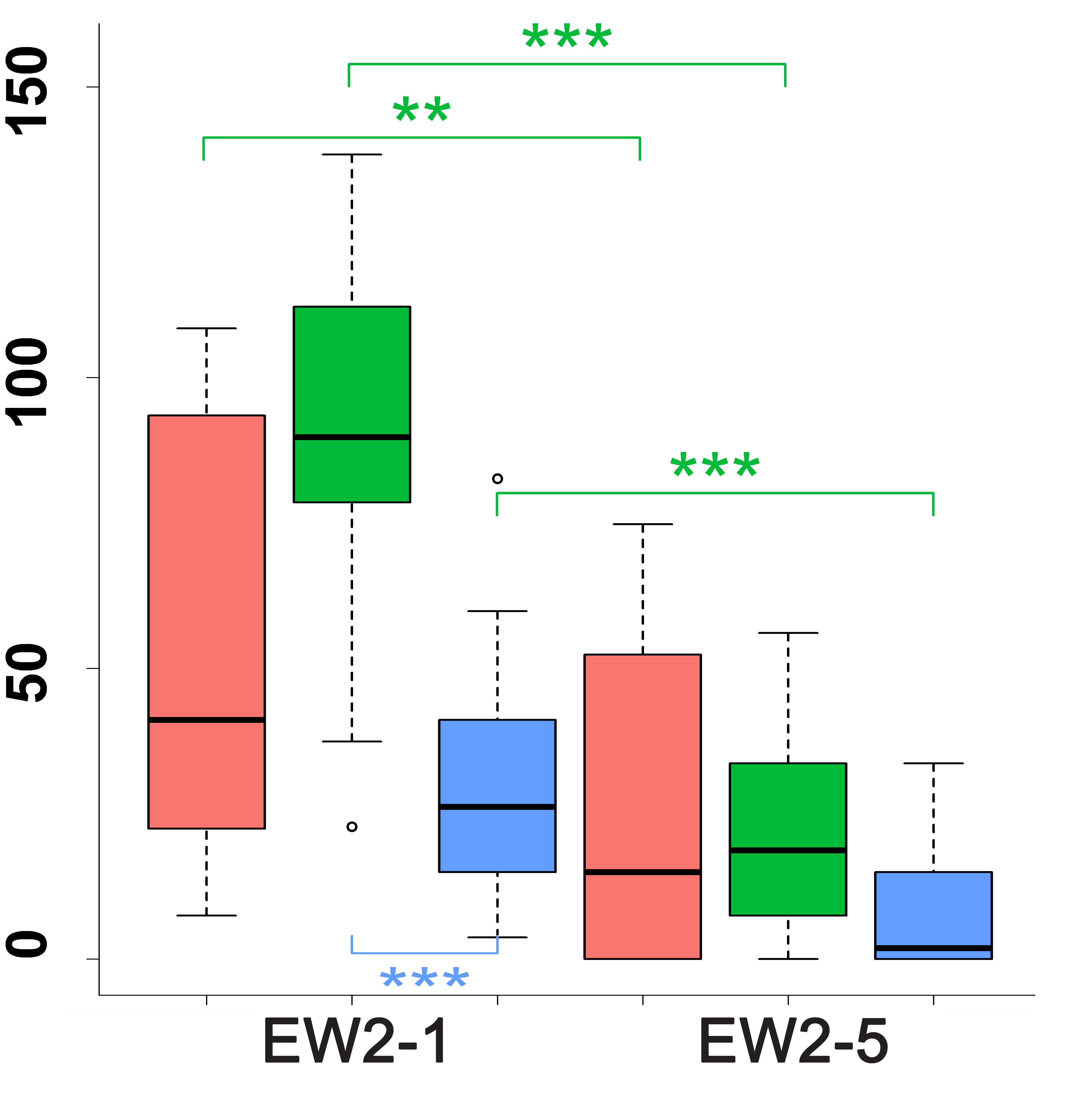}}
\hfill
\subfigure[EW1 Transfer SSQ]{\includegraphics[trim={0 2em 3em 0em}, clip, width=0.24\linewidth]{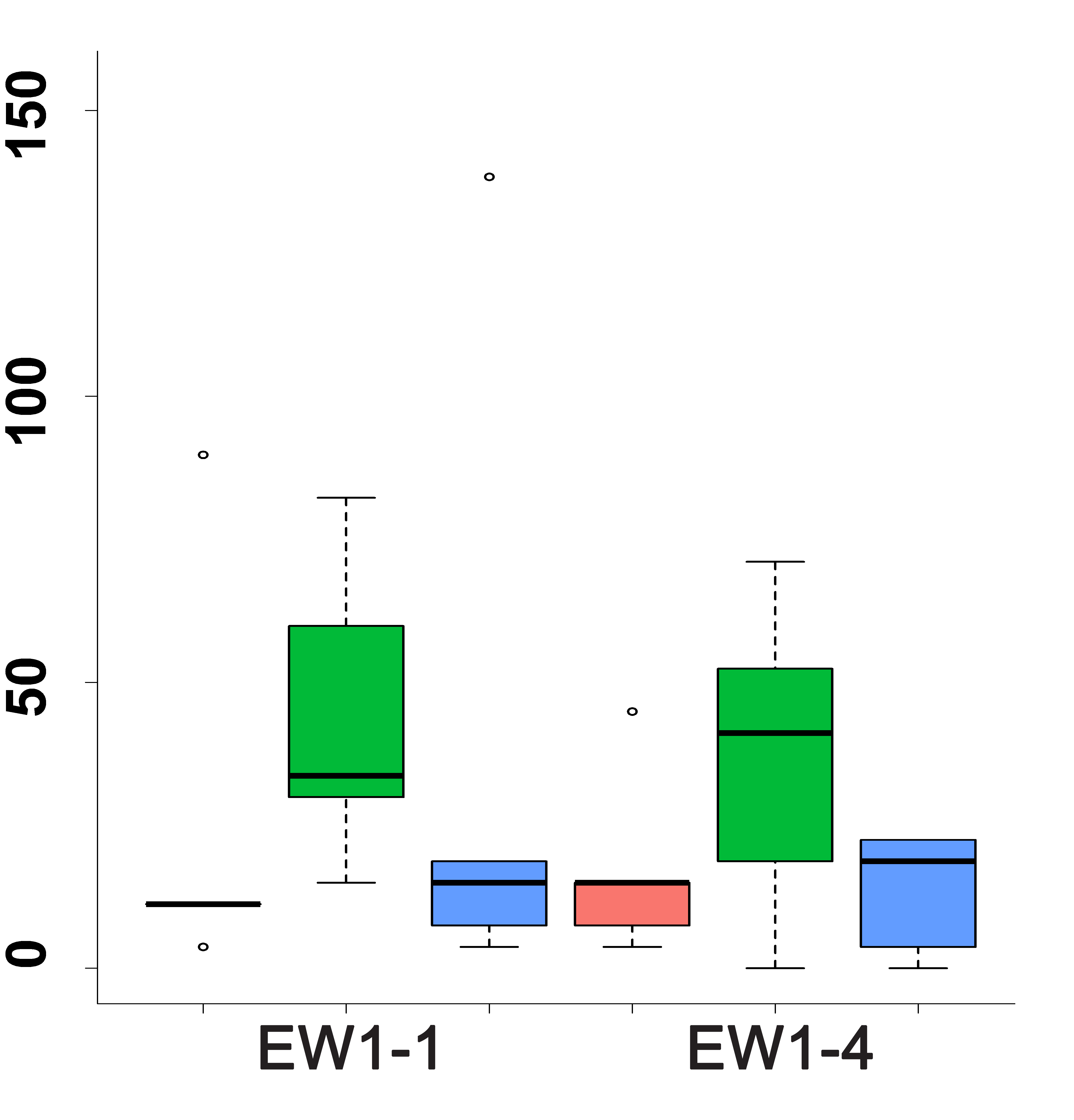}}
\hfill
\subfigure[EW2 Transfer SSQ]{\includegraphics[trim={0 2em 3em 0em}, clip, width=0.24\linewidth]{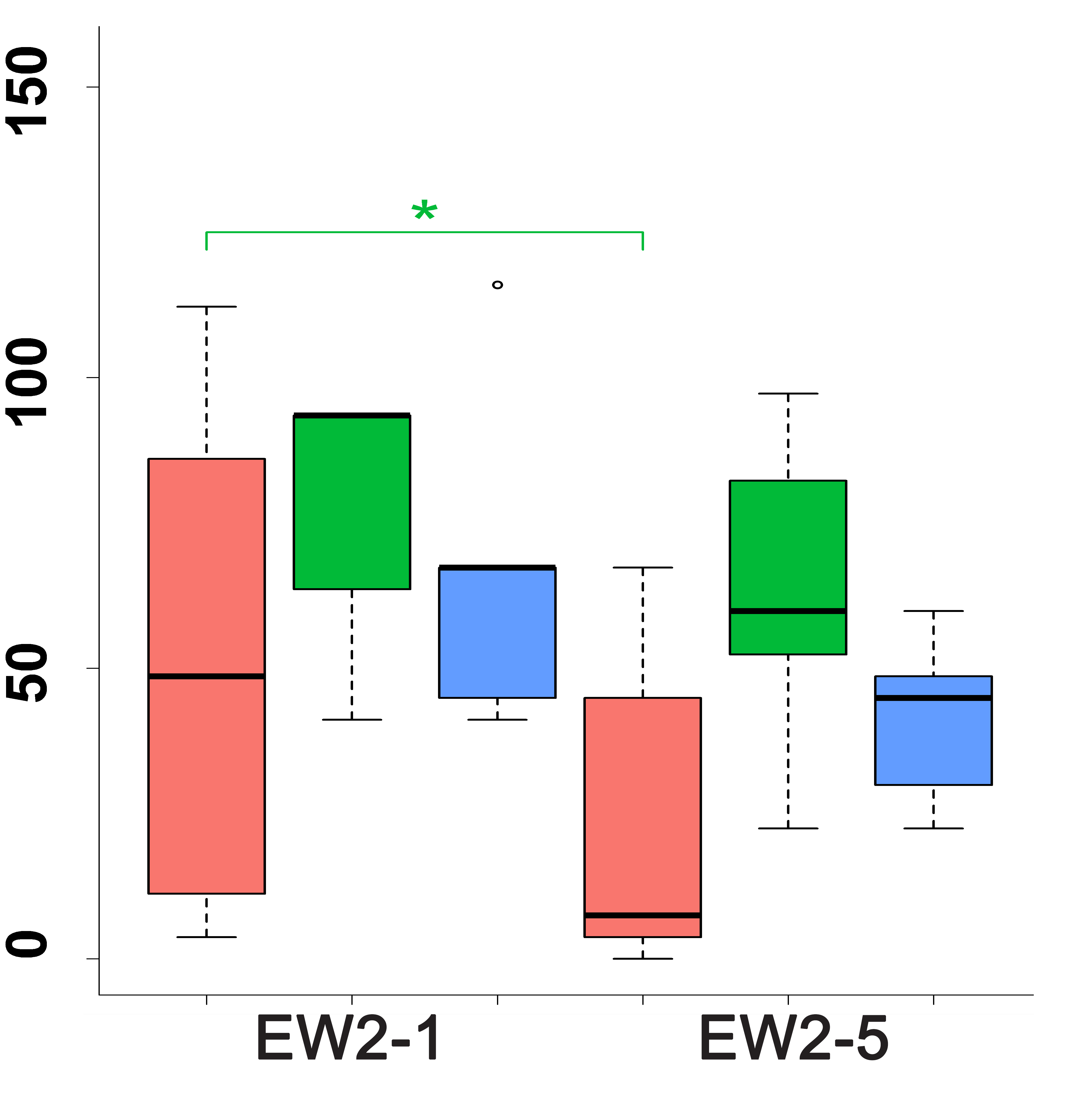}}
\caption{ 
% Boxplots of the 
Visual SSQ total scores for training content for VRT, VRO, and 2DT on the first and last days of EW1 and EW2 respectively (a, b), and those for the transfer content (c, d) which were tested on VR only
(* $p$\,\textless\,.05; ** $p$\,\textless\,.01; and *** $p$\,\textless\,.001). ~\label{nparLDananalysisResutls}
}
\end{figure*}

On the other hand, the pairwise comparisons with respect to the factor of
days (1 day vs. 4 day) show significant reductions in cybersickness levels for both VRT conditions (1 day $>$ 4 day; $p <$ .05) and 2DT conditions (1 day $>$ 4 day; $p <$ .05), that is, indicating the effectiveness of balance training methods (see Figure~\ref{nparLDananalysisResutls} (a) and Table~\ref{wilcoxon}). 
In contrast, there was no statistically significant decrease observed in the VRO condition.
It should be noted that the VRO condition solely involved exposure to visual stimulation (i.e., immersive VR viewing) without any balance training.
% There was no reduction observed in the VRO condition.
% Note that the VRO condition did not accompany any balance training but only involved exposure to visual stimulation, i.e., immersive VR viewing.

\subsubsection{EW2}

In EW2, unlike in EW1, 
significant effects were observed in both factors and their interaction: 
training methods ($p <$ .05), days ($p <$ .001), 
training methods $x$ days ($p <$ .05) (see Table~\ref{nparLDananalysisResutls}).
The pairwise comparison with respect to the training methods showed a statistically significant difference between VRO and 2DT on the first day, with VRO showing higher levels of cybersickness (VRO $>$ 2DT; $p <$ .05).
Due to the increased navigational complexity of the content used in EW2 compared to that of EW1 (see Figure~\ref{trajectories}), cybersickness levels increased overall in EW2, as expected. 
It is noteworthy that VRO, in particular, exhibited significantly higher levels compared to the training methods on the first day,
did not involve any prior balance training, nor did it show increased tolerance to sickness by the sustained exposure from EW1. 
Interestingly, the VRT group, which initially started with the highest cybersickness levels in EW1, demonstrated lower levels than VRO on the first day of EW2 (see Figure~\ref{nparLDananalysisResutls} (b)).
While these results hint at the effect of balance training on cybersickness (possibly supporting H2 and H4), similarly to EW1, effects analysis on the between-subject factor of the training method must be taken with a grain of salt.

As for the pairwise comparisons on the factor of days, we found statistical differences in all conditions, as follows: VRT (1 day $>$ 5 day; $p <$ .01), VRO (1 day $>$ 5 day; $p <$ .001), and 2DT (1 day $>$ 5 day; $p <$ .001), as shown in the Table~\ref{wilcoxon}.
In all conditions, the cybersickness scores were lower on the last day compared to the first day.
These findings strongly support H2. 
In addition, increased tolerance to cybersickness by simple sustained 
exposure and habituation to VR (VRO) is observed in EW2, while not so in EW1.  We posit that
its training effect necessitates a relatively longer time, only starting to take effect after EW2.

\subsection{Balance Performance Improvement}
\label{sec:balance performance improvement}

To relate the potential effect of balance training to increasing tolerance for cybersickness, various measures were taken over the course of the experiments (excluding the VRO subjects), such as (1) the duration of subjects' balance maintenance along with (2) the occurrences of balance failures (instances of subjects having to place their one foot back on the ground to avoid falling to the ground) and (3) the variability in their centers of mass.
The former was measured before and after the training sessions, while the latter two during. 

The comparative analysis with respect to the between-subject factor of the training method is not presented for the similar reason of the different subject groups of limited number.

% \textcolor{red}{Korea-hidden} \vsnote{}, 

\subsubsection{Before and after training}
% Balance performance improvement was also 
Figure~\ref{fig:Balance Performance} shows the balance performance trend during EW1 and EW2.
Balance maintenance was tested separately before and after EW1 and EW2 in the form of  the ``one leg stand with eyes closed'' test (see Figure~\ref{fig:process}).
This is a common balance performance test and the average performance for people in their twenties (demographics for our subject group) is reportedly about 20 seconds in certain countries~\cite{ART002391980}.
Note that this test was administered before and after the training sessions of EW1 and EW2 (one leg stand with eyes closed).

After confirming the normality test, the independent samples t-test was applied to examine for any differences between the two conditions (i.e., VRT and 2DT), however, no statistical differences were found among the tested days (see Figure~\ref{fig:Balance Performance} (a)).
However, when comparing the first day (before EW1) and the final day (after EW2), 
we observed a relatively large increase of approximately 27 seconds in the average balance maintenance time for the VRT condition, whereas the increase was only about 2 seconds for the 2DT condition. 
This suggests that the immersive environment (VRT) possibly had a greater impact on balance training compared to the 2D environment (2DT).
These findings partially support our third hypothesis (H3) that immersive balance training (VRT) can be more effective in improving balance compared to training in the 2D/non-immersive environment (2DT).

\subsubsection{During training}
\begin{figure}[h]
\centering
\subfigure[Average time of balance maintenance (one leg stand with eyes closed given before and after the training sessions of EW1 and EW2).]{\includegraphics[trim={0 10em 0 0}, clip, width=0.8\columnwidth]{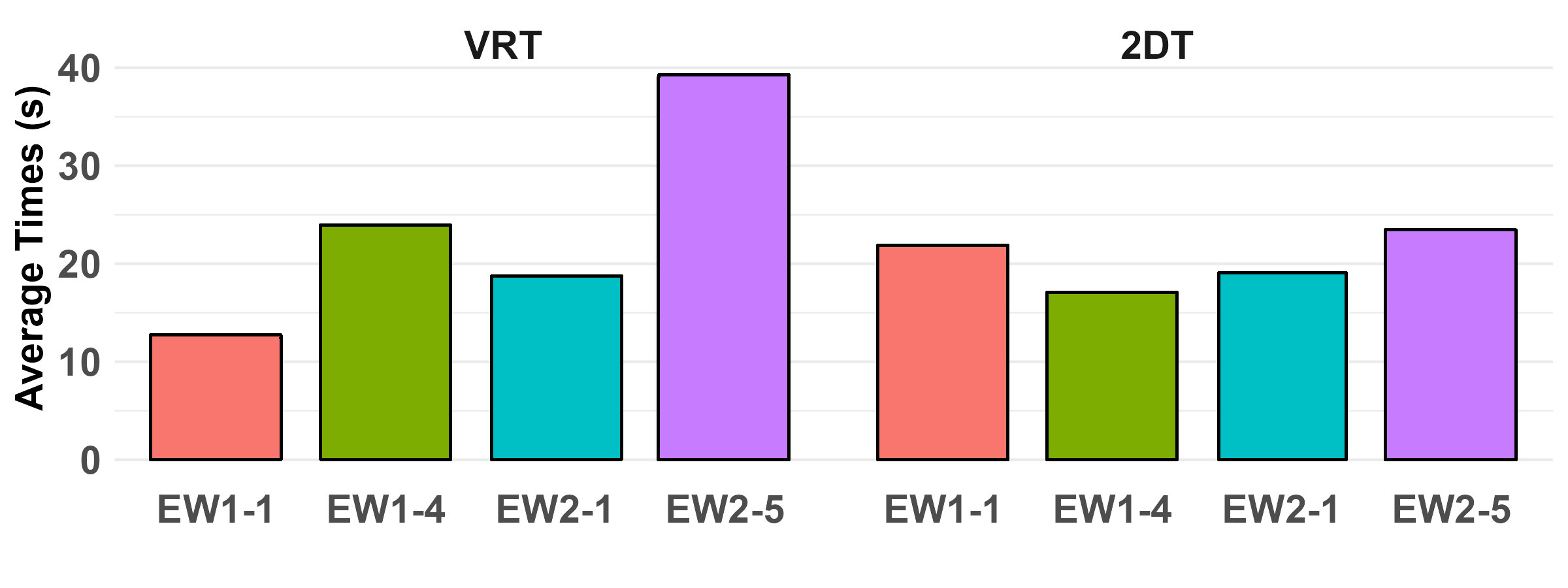}}
\hfill
\subfigure[Number of times of placing foot down (balance failures) measured during the training sessions of EW1 and EW2 (one leg stand while watching the training content).]{\includegraphics[trim={0 1em 0 0}, clip, width=0.8\columnwidth]{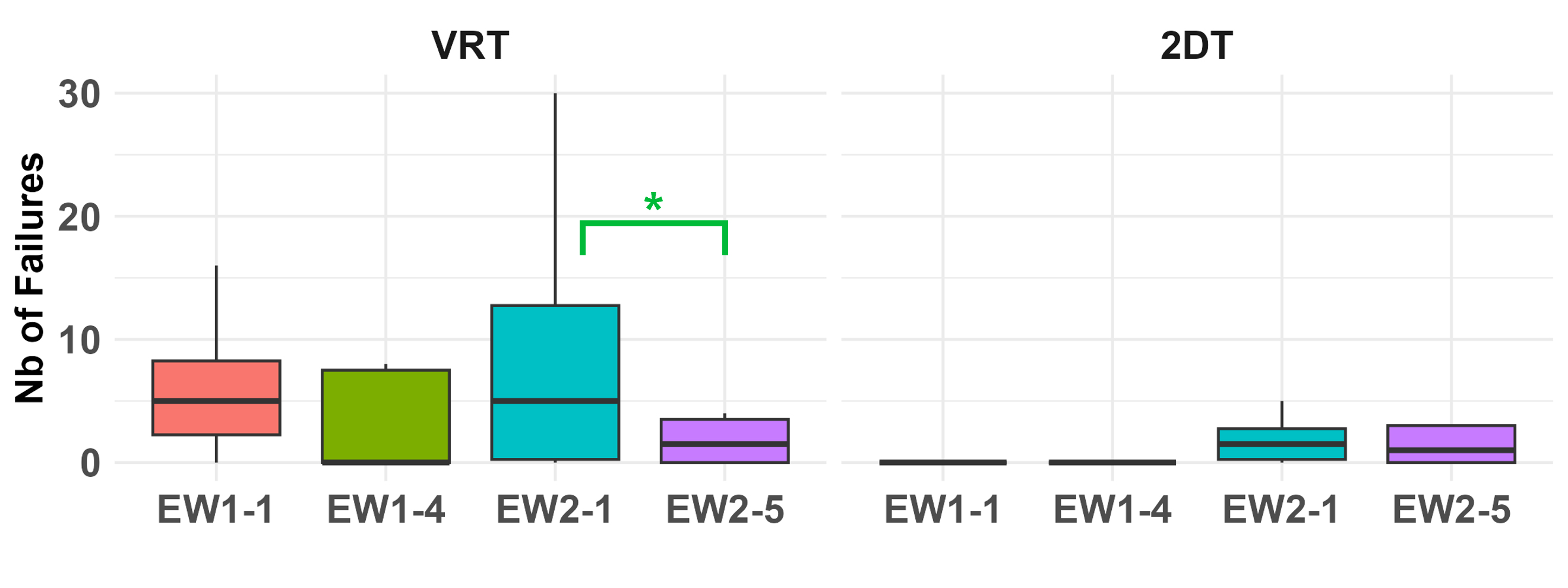}}
\subfigure[Center of mass variability measured during the training sessions of EW1 and EW2 (one leg stand while watching the training content). ]{\includegraphics[trim={0 1em 0 0}, clip, width=0.8\columnwidth]{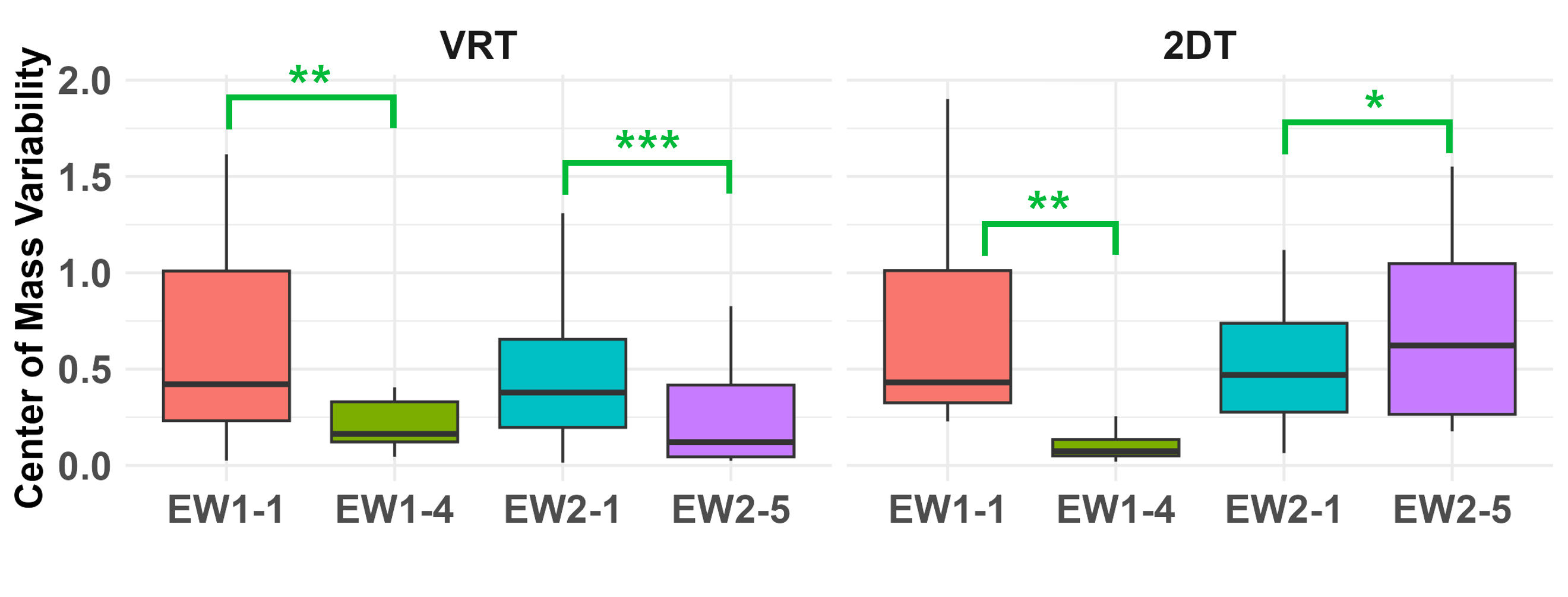}}
\caption{
Balance performance trend over EW1 and EW2 between VRT and 2DT: (a) Average time of balance maintenance (one leg stand with eyes closed before and after the EW1/EW2 training sessions), 
(b) Number of times of placing the foot down,
and (c) Center of mass variability (measured for one leg stand while watching the training contents).
\label{fig:Balance Performance}
}
\end{figure}

We obtained similar results for the number of balance failure cases and center of mass variability, as shown in Figure~\ref{fig:Balance Performance} (b) and (c) respectively.
Note that these quantities were measured during the training sessions of EW1 and EW2 while watching the training contents.
To analyze the data, we first assessed its normality. 
Next, we performed separate comparisons within each week, using the Wilcoxon sign rank test, between the initial and final days for both the VRT and 2DT conditions.
In the case of the number of failures, where participants put their feet on the ground, a significant effect was observed between EW2-1 and EW2-5 in the VRT condition ($p <$ .05; EW2-1 $>$ EW2-5). This finding strongly indicates that immersive VR environments can enhance balance performance.
% In the case of the number of failures, which put their feet on the ground, there is one significant effect between EW2-1 and EW2-5 in VRT condition ($p <$ 0.5; EW2-1 > EW2-5). 
% It means that immersive VR environments can enhance balance performance. 
%

The lack of significant differences observed in the first week may be attributed to the relatively easier task situation (i.e., relatively less sickness), as shown in Figure~\ref{fig:firstweek}. 
As a result, the number of balance failure cases was relatively lower.
Moreover, in the 2DT condition, for which the subjects did not wear a VR headset, 
the visible real environment (e.g., wall, floor) might have helped the subjects attain their balance as well.
On the contrary, despite the increased difficulty of the task/content in EW2, the training regimen resulted in a notable reduction in the number of failures.

The center of mass variability results are illustrated in Figure~\ref{fig:Balance Performance} (c), and there were significant differences in all combinations.
%Figure~\ref{fig:Balance Performance} (c) includes only variability values of 2 or less in the graph. 
%Values beyond this range were excluded from the figure as they represented a phase where subjects had to hop to stabilize their body's center of mass in order to prevent touching the ground or falling.
%This could clearly express the difference between conditions.
In the first week (EW1), both the VRT and 2DT groups significantly decreased in their variability by the 4th day (last day, EW1-4), compared to the first day (day 1, EW1-1) - 
VRT: $p <$ .01; 2DT: $p <$ .01.
However, during the second week (EW2), while there was a significant decrease in VRT (EW2-1 $>$ EW2-5; $p <$ .001), 2DT showed a significant increase (EW2-1 $<$ EW2-5; $p <$ .05).
Again, we believe this is due to the first week (EW1) environment/content 
being relatively monotonous, making it easier for them to maintain a stable center of balance, making it difficult for all the factors to exhibit any effect.

Overall, the data show the expected results of subjects' balance capability improving in time for VRT and 2DT. Note that VRO involved no balance training.  
With categorical statistical significance, it moderately supports our first hypothesis (H1).

\subsection{Correlation between balance and sickness}

To further investigate the relationship between balance performance and the reduction in ``Visual'' sickness, a correlation analysis was conducted using the Pearson correlation coefficient test. The following null hypothesis value of correlations were made: (1) ``Visual" sickness scores and balance maintenance time would be negatively correlated (-1); (2) ``Visual" sickness and the number of balance fails would be positively correlated (+1); and (3) ``Visual" sickness and center of mass variability would be positively correlated.
The results are given in Table~\ref{correlation} and they are mostly consistent with our hypotheses (e.g., H2 and H3).

Statistically significant correlations were found between the 
improvements in the number of balance failures and center of mass variability, respectively (either by VRT or 2DT) with the ``Visual'' sickness scores over EW1 and EW2.
\note{VRT showed higher correlation coefficients than the 2DT in two measures}
(i.e., No. of balance failures: $r = 0.612 > 0.267$; center of mass variability: $r = 0.305 > 0.269$).
For the balance maintenance time, 
there was no significant correlation found, however,
it is worth noting that while VRT showed a negative correlation as expected ($r = -0.295$), 2DT only showed 
near zero correlation ($r = 0.001$).

To summarize, as the balance performance improves, 
there is an increase in tolerance to the ``Visual'' sickness in both immersive (VRT) and non-immersive (2DT) environments.  Furthermore, the correlation values 
indicate that the training effect in the immersive VR environment (VRT) was more pronounced than that in the non-VR (2DT).

\begin{table}[h]
\caption{The Pearson correlation analysis between ``Visual'' sickness scores and balance performances (* $p$\,\textless\,.05; ** $p$\,\textless\,.01; and *** $p$\,\textless\,.001)~\label{correlation}}
\renewcommand{\arraystretch}{1.3}
\centering
\begin{tabular}{lcc}
\hline \hline
& \multicolumn{2}{c}{\textbf{Visual SSQ}}
\\
\textbf{Balance Performance }
& \textbf{VRT }
& \textbf{ 2DT}
\\ \midrule
\textbf{Maintenance time}
& $r =$ -0.295; $p =$ .103\hphantom{***}\hphantom{<} 
& $r =$ 0.001; $p =$ .501\,\,  \\ 
\textbf{Number of balance failures}
& $r =$ \hphantom{-}0.612; $p <$ .001 ***  
& $r =$ 0.267; $p =$ .048 *  \\ 
\textbf{Center of mass variability}
& $r =$ \hphantom{-}0.305; $p =$ .028 *\hphantom{<}\hphantom{**}   
& $r =$ 0.269; $p =$ .049 *  \\ 
\hline \hline
\end{tabular}
% }
\end{table}

\subsection{Transfer Effect}
\label{sec:transferEffect}
The true test for any effect of the balance training on cybersickness would be 
shown by observing how the balance-trained subjects perform on completely different VR contents,
namely, the ``transfer'' contents 
that were totally different from the test content (see Figure~\ref{transfereffectvirtualenv}).  
The assessments were made two times, before and after EW1 using transfer VR content 1 (Rollercoaster ride), 
and before and after EW2 using transfer VR content 2 (Space navigation).  
The sickness levels were measured in the same way as those conducted during the training sessions of EW1 and EW2.
The results are summarized in Figure~\ref{nparLDananalysisResutls}, 
Table~\ref{table:nparLD}, Table~\ref{kruskal}, and Table~\ref{wilcoxon}.

The statistical analysis (see Table~\ref{wilcoxon} and Figure~\ref{nparLDananalysisResutls} (d)) found a significant decrease in the cybersickness 
only in the VRT condition ($p <$ .05) before and after EW2.
No other condition exhibited any statistically significant reduction (i.e., VRO, 2DT).
This indicates the training transfer effect of balance training in VR for cybersickness tolerance, consistent with 
the observation that the VRO group, who received no training in EW1, showed much higher levels
of sickness in the early stages of EW2 (when switched to the new training content) than VRT.   
Furthermore, it confirms our hypotheses (H3 and H4) that immersive media (VRT) with balance training is a more effective method than simply being
exposed to the VR content for an equal amount of time.

%% file: sections/06_Discussion.tex
% \begin{table}[h!]
%   \centering
%   \caption{Results of 2x3 Repeated Measures ANOVA with Multiple Dependent Variables}
%   \begin{tabular}{@{} lcccccc @{}}
%     \toprule
%     & \multicolumn{3}{c}{Dependent Variable 1} & \multicolumn{3}{c}{Dependent Variable 2} \\
%     \cmidrule(lr){2-4} \cmidrule(lr){5-7}
%     Factor & F-value & p-value & \( \eta_p^2 \) & F-value & p-value & \( \eta_p^2 \) \\ 
%     \midrule
%     Time (2 levels) & [F] & [p] & [η²] & [F] & [p] & [η²] \\
%     Training Method (3 levels) & [F] & [p] & [η²] & [F] & [p] & [η²] \\
%     Time x Training Method & [F] & [p] & [η²] & [F] & [p] & [η²] \\
%     % Repeat rows for additional dependent variables if needed
%     \bottomrule
%   \end{tabular}
%   \label{tab:my_label}
% \end{table}

\section{Discussion}

\subsection{The Effect of Balance Training on Cybersickness}

As discussed in length in Section~\ref{sec:cybersickness}, there have been several
theories as to why and how cybersickness occurs, such as the sensory mismatch~\cite{laviola2000discussion, Rebenitsch}, 
lack of the rest frame~\cite{Harm}, 
and postural instability~\cite{Riccio, Ruixuan, Smart}.  
All such factors are plausible and debatable at the same time.
While the proposal of immersive balance training for developing tolerance to cybersickness hinges on the postural instability in particular, 
it does not discount the effect of those other factors nor is it in conflict with them. 

%we consider that the precise cause of cybersickness is still unknown, and thus treat all theories equally, without prioritizing any one theory over the others. 
%Despite the specific focus of our research, this section discusses the effects of balance training on cybersickness and an extended understanding based on its relationship with other theories and alternative solutions.

% As described in Section~\ref{sec:results}, significantly reducing trends were observed in all treatments.

% As described in Section~\ref{sec:results}, 

Our experiments have shown significant reductions in cybersickness symptoms in all the treatments. 
This trend was also observed, albeit to a lesser extent, in the VRO treatment that did not include training. 
These results indicate that repeated exposure to VR contents reduced sickness~\cite{adhanom2022vr, palmisano2022reductions}, and it is difficult to deny that this effect may have influenced other conditions as well (in terms of what contributed to the reduction).
On the other side, balance training may have reduced the sickness acting
as cognitive distraction~\cite{kourtesis2023cybersickness, venkatakrishnan2023cog02}.
However, cognitive distraction alone is difficult to explain the training transfer effect.
The same is true as for the mere exposure to a particular VR content.

% 이러한 결과는 VR에 반복된 노출이 sickness를 줄였음을 의미하며, 다른 조건에서도 이에 대한 영향이 포함되었음을 부정하는 것은 어렵다. 
%

Palmisano et al.~\cite{palmisano2022reductions} have shown
that repeated exposure to VR contents could significantly improve cybersickness.
However, this improvement was observed only for the very content the subjects were
exposed to, and it was not shown whether the effect extended to other 
VR contents.
On the other hand, our experiment confirmed the transfer effect of the balance training 
to a completely different content.
Only the VRT group, which engaged in the immersive balance training, significantly saw
the reduction in cybersickness in the transfer contents (see Figure~\ref{nparLDananalysisResutls}).  
This is the critical finding that sets forth the training (and the improved physical/mental capability)
as the main culprit to the sickness reduction - more so than the exposure itself or distraction.
This also signifies the potential practicality of the approach.

\subsection{The Potential of Balance Training on Cybersickness}

One representative experiment in the attempt to validate the postural instability theory by~\cite{Riccio} showed the decreasing sickness levels in a provocative situation by the subject making a more widened and stable stance~\cite{Dennison}.  
In contrast, the experiment in this work went to other ways, where the subjects were purposely situated to be unstable (one leg stand), leading to a possible expectation that the ``sickness'' should increase according to the same theory.
One important difference, however, is that the subjects were also instructed to ``learn'' and train as to how to maintain the balance.  
Indeed, instead of the increased level of sickness, our results clearly show the reduction and even the transfer effects, singling out the very effect of the ``training''.

Interestingly, according to Menshikova et al.~\cite{Menshikova}, when compared figure skaters, soccer players, and wushu fighters, figure skaters showed the most resilience to cybersickness. 
Thus, innate or learned balancing capability seems related to tolerance to cybersickness.  
Ritter et al.~\cite{Ritter} studied the VR-based (safe) training of balance beam performance with gymnastics beginners.  
Among others, the work showed that the subjects generally performed worse in VR than in the real world.
This indirectly suggests that, for cybersickness improvement by balance training, the training environment (i.e., VR) will be important.  
Likewise, our results point similarly in the same direction, namely, VRT being more effective than 2DT and even VRO.  

As for 2DT, the level of the sickness arising from the visual motion must have been less so to begin with compared to that by VR. 
The visual content has a substantially smaller field of view (approximately VRT: 100$^{\circ}$  vs. 2DT: 60$^{\circ}$) outside which objects possibly acting as reference objects are visible (e.g., walls, floor). 
These are aspects that can diminish the training effect in 2DT as well.  
On a similar note, training for a spatial task (which the balancing or even withstanding cybersickness from visual motion could be examples of) on the 2D oriented desktop environment has shown a negative transfer effect to the corresponding 3D VR environment~\cite{Pausch}.

Even though our study seems to show that extended exposure to VR does have an effect on building tolerance to cybersickness, in relation to the related work (see Section 2), its firm establishment is still debatable. Even if it was, we believe that its effect is weaker and not so long-lasting than that of balance training.  
In balance training, the user makes a conscious effort to encode the relevant information into one's proprioceptive and muscular control system.  
How long the training effect can be sustained would be a topic of future research.

\subsection{Limitations and Future Works}

Our study is limited in several aspects.  
Cybersickness is a truly multifactorial issue, including gender, age, the nature of the tasks undertaken, type of feedback, and multimodality~\cite{feng2016tactile, peng2020walkingvibe}
and the types of devices used~\cite{kourtesis2023cybersickness, kim2008application, chang2020virtual}. 
Our work only investigated one such probable factor, i.e., balancing capability.
While most factors mentioned above are known to influence the level of cybersickness in one way or another, the variance from the individual difference is relatively large~\cite{tian2022review, chang2020virtual, howard2021meta}.
Balancing capability can be considered a more predictable control factor~\cite{ARCIONI20193,doi:10.1080/10447318.2017.1286767}.
Training for it is also expected to be much less dependent on the immersive training environment (content genre).
% 디바이스나 컨텐츠에 상관없이 사람들은 조금의 balance capacitiy를 가지고 있고, 그것이 디바이스나 컨텐츠에 상관없이 조금이라도 증가할 것이다. 라는 의미 일건데, 
% 지금은 약간 다르게 해석될 여지가 있는 듯 합니다. 
% \note{citation needed}
Note that the training process can be further expedited by employing multimodal feedback, guidance features, and gamification~\cite{10.1145/3562939.3567818,Juras,Prasertsakul}.
Such are subjects of future research topics.

%while an individual's physical balance abilities may differ widely, they are intrinsic abilities that all individuals possess\textcolor{red}{[citation needed]}.
%Regardless of gender or age, these capabilities can be improved through training and are not altered by the presence or absence of a task, or by differences in equipment. 

%In addition, since it is an intrinsic factor, there is no need to prepare multimodal feedback, or external factors, separately to reduce the sensor gap.
%Thus, we believe enhancing physical balance through training represents a universal strategy that could potentially alleviate cybersickness.}

%This may not be simplistically attributed to a single cause, such as the imbalance we explored.
%Furthermore, our study did not include multimodal feedback to mitigate sensory gaps, which might have led to more severe cybersickness~\cite{feng2016tactile, peng2020walkingvibe}.}

% 개인의 균형 능력은 차이가 있지만 개인들이 갖는 공통의 능력이다. 성별, 나이와 관계없이 훈련을 통해 개량할 수 있고, task가 있거나 없거나 혹은 기기의 차이에 의해서도 변화되지 않는 개인의 능력이다. 따라서, 우리는 balance training을 통한 신체 능력 향상이 VR 멀미를 극복할 수 있는 잠재력은 가진 기초적인 방법 중 하나라고 고려한다.
% 

% \textcolor{red}{need to mention multi modal feedback}

Another limitation is the number of experimental subjects in each of the training method (between-subject) groups. 
The subjects were also confined to a particular group, i.e., young adult males.  
It is too early to generalize our claims to other subject populations. 
%
%Thus, aside from the results from the transfer content, we reserved the direct comparison with respect to the training method.  
A future larger-scale experiment should not only accommodate a larger number of subjects but also employ a variety of sickness-eliciting or ``provocative'' contents as well. 
As there may be more fitting and proper balance training routines, these new VR contents may involve interaction techniques to guide such balancing acts more effectively as demonstrated in~\cite{Yang}.

%% file: sections/07_Conclusion.tex
\section{Conclusion}
In this paper, we conducted a two-week-long experiment to observe the relationship between user balance learning and developing sickness tolerance 
under different experimental conditions.
The findings indicate that enhancing individual balance performance leads to an increased tolerance for cybersickness. 
The study also corroborated for the greater effectiveness of balance training in immersive environments compared to non-immersive settings.
Furthermore, the improvement in the balance ability demonstrated sustainable effects, enabling individuals to tolerate VR motion sickness in newly encountered VR environments as well.

Although our results are still preliminary, it is the first of its kind.  If further validated with continued in-depth and larger scale studies, we hope to be able to design and recommend a standard VR-based balance training regimen for building tolerance to sickness for active yet sickness-sensitive ``wannabe'' VR users (while also improving one's fitness at the same time as a bonus).

%% file: sections/A0_Appendix.tex
\appendix

\section{Questionnaire}

\begin{table}[h]
\renewcommand{\arraystretch}{1.2}
% \centering
\caption{Overall SSQ questionnaire}
\resizebox{1\linewidth}{!}{
\begin{tabular}{  p{0.05\linewidth} | p{1.3\linewidth}  }
    \hline \hline
    Q01 & I felt uncomfortable while experiencing the content. \\ \hline
    Q02 & I felt fatigued while experiencing the content. \\ \hline
    Q03 & I felt a headache while experiencing the content. \\ \hline
    Q04 & I felt eye strain while experiencing the content. \\ \hline
    Q05 & I found it difficult to keep my eyes focused while experiencing the content. \\ \hline
    Q06 & I felt an increased amount of salivation while experiencing the content. \\ \hline
    Q07 & I felt nervous and sweaty while experiencing the content. \\ \hline
    Q08 & I felt nausea while experiencing the content. \\ \hline
    Q09 & I found it difficult to concentrate while experiencing the content. \\ \hline
    Q10 & I experienced a head full feeling while performing the content. \\ \hline
    Q11 & I experienced a feeling of blurred vision while performing the content. \\ \hline
    Q12 & I felt dizzy when I opened my eyes after experiencing the content. \\ \hline
    Q13 & I felt dizzy when I closed my eyes after experiencing the content. \\ \hline
    Q14 & I felt vertigo while experiencing the content. \\ \hline
    Q15 & I felt a stomach awareness, experiencing the content. \\ \hline
    Q16 & I felt burping while experiencing the content. \\ \hline \hline
\end{tabular}
}
\end{table}

\begin{table}[h]
    \renewcommand{\arraystretch}{1.2}
    % \centering
    \caption{Visual SSQ questionnaire}
    \resizebox{1\linewidth}{!}{
    \begin{tabular}{  p{0.05\linewidth} | p{1.3\linewidth}  }
        \hline \hline
        Q01 & I felt uncomfortable, particularly by the visual content and stimulation.   \\ \hline
        Q02 & I felt fatigued, particularly by the visual content and stimulation.   \\ \hline
        Q03 & I felt a headache, particularly by the visual content and stimulation.        \\ \hline
        Q04 & I felt eye strain, particularly by the visual content and stimulation.                \\ \hline
        Q05 & I found it difficult to keep my eyes focused, particularly by the visual content and stimulation.\\ \hline
        Q06 & I felt an increased amount of salivation, particularly by the visual content and stimulation.   \\ \hline
        Q07 & I felt nervous and sweaty, particularly by the visual content and stimulation.               \\ \hline
        Q08 & I felt nausea, particularly by the visual content and stimulation.\\ \hline
        Q09 & I found it difficult to concentrate particularly by the visual content and stimulation.         \\ \hline
        Q10 & I experienced a head full feeling, particularly by the visual content and stimulation.          \\ \hline
        Q11 & I experienced a feeling of blurred vision, particularly by the visual content and stimulation.  \\ \hline
        Q12 & I felt dizzy when I opened my eyes after experiencing the content, particularly by the visual content and stimulation.    \\ \hline      
        Q13 & I felt dizzy when I closed my eyes after experiencing the content, particularly by the visual content and stimulation. \\ \hline
        Q14 & I felt vertigo, particularly by the visual content and stimulation.\\ \hline
        Q15 & I felt a stomach awareness, particularly by the visual content and stimulation.   \\ \hline
        Q16 & I felt burping due to the visual cues provided.         \\ \hline
        \hline
    \end{tabular}
    }
\end{table}

\begin{table}[h]
    \renewcommand{\arraystretch}{1.2}
    % \centering
    \caption{Balance SSQ questionnaire}
    \resizebox{1\linewidth}{!}{
    \begin{tabular}{  p{0.05\linewidth} | p{1.3\linewidth}  }
        \hline \hline
        Q01 & I felt uncomfortable particularly due to the required stance and/or balancing routine.         \\ \hline
        Q02 & I felt fatigue due to the required stance and/or balancing routine.       \\ \hline
        Q03 & I felt a headaches due to the required stance and/or balancing routine.      \\ \hline
        Q04 & I felt eye strain due to the required stance and/or balancing routine.       \\ \hline
        Q05 & I found it difficult to keep my eyes focused due to the required stance and/or balancing routine.          \\ \hline
        Q06 & I felt an increased amount of salivation due to the required stance and/or balancing routine.   \\ \hline
        Q07 & I felt nervous and sweaty due to the required stance and/or balancing routine.   \\ \hline
        Q08 & I felt nausea due to the required stance and/or balancing routine.     \\ \hline
        Q09 & I  found it difficult to concentrate due to the required stance and/or balancing routine.   \\ \hline
        Q10 & I experienced a head full feeling due to the required stance and/or balancing routine.   \\ \hline
        Q11 & I experienced a feeling of blurred vision due to the required stance and/or balancing routine.  \\ \hline
        Q12 & I felt dizzy when I opened my eyes after experiencing the content due to the required stance and/or balancing routine.      \\ \hline
        Q13 & I felt dizzy when I closed my eyes after experiencing the content due to the required stance and/or balancing routine.   \\ \hline
        Q14 & I felt vertigo due to the required stance and/or balancing routine.     \\ \hline
        Q15 & I felt a stomach awareness due to the required stance and/or balancing routine.   \\ \hline     
        Q16 & I felt burping due to the required stance and/or balancing routine.       \\ \hline
        \hline
    \end{tabular}
    }
\end{table}